# Multi-Modal Multi-Agent System for Bridge Operation and Maintenance based on Large Language Model

Xiangyun Luo [a, b], Yunkai Zhang [a, b], Yong Huang [a, b,*], Stephen Wu [c, d], Hui Li [a, b]

a Key Lab of Smart Prevention and Mitigation of Civil Engineering Disasters of the Ministry of Industry and Information Technology, Harbin Institute of Technology, Harbin 150090, China

b Key Lab of Structures Dynamic Behavior and Control of the Ministry of Education, School of Civil Engineering, Harbin Institute of Technology, Harbin 150090, China

c The Institute of Statistical Mathematics, Research Organization of Information and Systems, 10-3 Midori-cho, Tachikawa, Tokyo 190-8562, Japan

d The Graduate University for Advanced Studies, SOKENDAI, 10-3 Midori-cho, Tachikawa, Tokyo 190-8562, Japan

**Correspondence**

Yong Huang

Email: huangyong@hit.edu.cn

**Abstract**

Bridges are critical transportation infrastructure, and reliable operation and maintenance (O&M) is essential for traffic safety and long-term structural performance. However, bridge O&M faces key bottlenecks: no unified architecture coordinates heterogeneous resources; model inferences lack clear links to physical–mechanical laws; and experts' tacit knowledge remains difficult to formalize. This study proposes a knowledge-enhanced autonomy formalization methodology and apply it to construct the Bridge O&M Multi-Modal Multi-Agent System (BOMMA), which integrates heterogeneous diagnostic actions, physics-mechanics-informed representations, long-term memory, and structured experiential knowledge within a coordinated multi-agent workflow, enabling interpretable and traceable scheduling, analysis, and decision-making. On 3,492 multi-source instructions, the proposed method achieves scheduling recalls of 85.37% for simple and 99.40% for composite tasks, with 95.35% BERT F1-score for O&M action reasoning. Ablation studies confirm the contributions of the components in the autonomy. This study bridges general LLM reasoning and specialized bridge O&M requirements.

## Nomenclature

Table 1 Symbols about the Agent's Reasoning Process

| Symbol | Indication |
|---|---|
| $\boldsymbol{X}$ | A token sequence generated by the agent. |
| $\boldsymbol{X}_A$ | The token sequence of the actions scheduled according to O&M instructions. |
| $\boldsymbol{X}_A^{(t)}$ | The token sequence of the t-th action in $\boldsymbol{X}_A$. |
| $\boldsymbol{X}_{\text{Dec}}$ | The token sequence of O&M decision. |
| $\boldsymbol{X}_I$ | The token sequence of an O&M instruction. |
| $\boldsymbol{X}_M$ | The token sequence of the agent's memory. |
| $\boldsymbol{X}_O$ | The aggregated token sequence of the action outputs. |
| $\boldsymbol{X}_O^{(t)}$ | The token sequence of the outputs of $\boldsymbol{X}_A^{(t)}$. |
| $\boldsymbol{X}_{\text{PM}}$ | The token sequence of the physical-mechanical principles of the action space. |
| $\boldsymbol{X}_{\text{PM}}^{(t)}$ | The token sequence of the part of the physical-mechanical principles about $\boldsymbol{X}_A^{(t)}$. |
| $\boldsymbol{X}_{\text{Pr}}$ | The token sequence of prompts for LLMs. |
| $\boldsymbol{X}_q^{(k)}$ | The token sequence of the k-th unstructured expert knowledge semantic part. |
| $\boldsymbol{X}_R$ | The aggregated token sequence of the action reasoning. |
| $\boldsymbol{X}_R^{(t)}$ | The token sequence of the action reasoning of action $\boldsymbol{X}_A^{(t)}$. |
| $\boldsymbol{X}_{vrv}^{(k)}$ | The token sequence of the k-th group of graph tuple. |
| $x^{(1):(i-1)}$ | The former i-1 tokens of $\boldsymbol{X}$. |
| $x_R^{(t),(j)}$ | The j-th token of the t-th action reasoning sequence $\boldsymbol{X}_R^{(t)}$. |
| $x_R^{(t),(1):(j-1)}$ | The former j-1 tokens of $\boldsymbol{X}_R^{(t)}$. |
| $\mathcal{G}$ | O&M expert knowledge graph. |
| $\mathcal{G}_R$ | The subgraph retrieved from $\mathcal{G}$. |
| $\mathcal{V}_I$ | A union of graph nodes whose relevance exceed the threshold $\lambda$. |
| $\mathcal{V}_R$ | A union of graph nodes which are in L-hop connection with $\mathcal{V}_I$. |
| $v^{(i)}$ | A graph node in O&M expert knowledge graph $\mathcal{G}$. |
| $v_o^{(k)}$ | The object node in the k-th semantic part of unstructured knowledge $\boldsymbol{X}_q^{(k)}$. |
| $v_s^{(k)}$ | The subject node in the k-th semantic part of unstructured knowledge $\boldsymbol{X}_q^{(k)}$. |
| $\mathcal{R}_R$ | The relation net of the retrieved subgraph $\mathcal{G}_R$. |
| $r^{(k)}$ | The relation in the k-th semantic part of unstructured knowledge. |
| $p(\bullet)$ | A conditional probability distribution predicted by LLMs. |

Table 2 Symbols about the Evaluation metrics for BOMMA's Experiments

| Symbol | Indication |
|---|---|
| Rec | Scheduling recall rate. |
| C-Pre($\boldsymbol{X}_I$) | Scheduling precision rate of composite instruction $\boldsymbol{X}_I$. |
| C-Rec($\boldsymbol{X}_I$) | Scheduling recall rate of composite instruction $\boldsymbol{X}_I$. |
| ExactMatch | The ratio of composite instructions that necessarily and sufficiently fulfill their intended scheduling without further interventions. |
| Itv | Average human intervention count. |
| R-Time | Average relative task time. |
| $P_{BERT}$ | BERT precision. |
| $R_{BERT}$ | BERT recall. |
| $F1_{BERT}$ | BERT F1-score. |
| $n_{\mathrm{TP}}$ | The number of simple instructions that successfully trigger the expected actions. |
| $n_{\mathrm{FN}}$ | The number of simple instructions that fail to trigger the expected actions. |
| $n_{\mathrm{Per}}$ | The number of testing O&M composite instructions that necessarily and sufficiently fulfill their intended scheduling without further interventions. |
| $n_{\mathrm{TC}}$ | The total number of testing composite O&M instructions. |
| $n_{\mathrm{Com}}$ | The number of zero-human-intervention composite instructions achieving their intended purpose under the complete BOMMA architecture. |
| $n_{\mathrm{Inc}}$ | The number of zero-human-intervention composite instructions achieving their intended purpose under certain types of incomplete BOMMA architecture. |
| $\boldsymbol{X}_{A(\mathrm{GT})}$ | The action scheduling ground truth of composite instruction $\boldsymbol{X}_I$. |
| $\boldsymbol{X}_{A(\mathrm{GT})}^{(j)}$ | One of the expected actions in $\boldsymbol{X}_{A(\mathrm{GT})}$. |
| $\boldsymbol{X}_{\mathrm{GT}}$ | The token sequences reasoned manually. |
| $x_{\mathrm{GT}}^{(j)}$ | An individual token in $\boldsymbol{X}_{\mathrm{GT}}$. |
| $rd^{(i)}$ | The number of additional human interventions required to fulfill the intended purpose of composite instruction $i$. |
| $tm'^{(i)}$ | The exact time for accomplishing instruction $i$ under certain types of incomplete BOMMA architecture. |
| $tm^{(j)}$ | The exact time for accomplishing instruction $j$ under the complete BOMMA architecture. |

## 1 Introduction

As a critical node of transportation infrastructure, the service condition of bridges directly relates to the operational safety and traffic efficiency of regional transportation networks. With the acceleration of bridge structural aging, the growth of traffic loads, and the frequent occurrence of extreme climate events, the pressure and complexity faced by bridge operation and maintenance (O&M) management have grown exponentially [1-3]. To achieve efficient, interpretable, and continuously learnable autonomous operation and maintenance decision-making for complex engineering scenarios, it is imperative to systematically integrate multi-source heterogeneous data, senior expert heuristics, and underlying physical and mechanical mechanisms within a unified framework, thereby establishing a traceable and trustworthy complete reasoning chain. There is thus an urgent need for a computational formalism that can represent heterogeneous engineering knowledge in executable, evidence-bearing, and auditable forms while enabling collaborative reasoning across modalities and O&M tasks.

The widespread introduction of artificial intelligence methods has significantly promoted the development of intelligent bridge O&M technology. In vehicle load identification, Zhang et al. [4] proposed a HardNet-based deep neural network for vehicle re-identification across bridge monitoring cameras. Deng et al. [5] introduced a physics-informed neural network for bridge influence surface identification, enabling BWIM with transverse load localization without prior axle knowledge. In damage diagnosis, Zhang et al. [6] proposed SDIGLM, a multi-modal chain-of-thought framework integrating VisualGLM-6B with U-Net-based visual reasoning for multi-type structural damage identification. Wei et al. [7] developed an end-to-end generative VRNN with a physically disentangled latent space for structural health condition diagnosis via reconstruction residuals. Deng et al. [8] proposed a RoMS-based damage identification method exploiting the correlation of global dynamic responses in high-dimensional space, which is mode-related and insensitive to traffic flow. In condition assessment and early warning, Qu et al. [9] proposed an MVO-CNN-BiLSTM model with mixed attention for bridge deformation quantile prediction, achieving interval prediction superior to point prediction. Ju et al. [10] developed an ATT-BiGRU fusion model with GMM/AIC-based dynamic thresholds for deflection early warning. In maintenance decision-making, Wei et al. [11] proposed a deep reinforcement learning framework based on Deep Q-Networks for optimal structural maintenance policy. Zhang et al. [12] introduced a multi-agent ranking proximal policy optimization (MARPPO) framework for bridge network life-cycle maintenance. Although the aforementioned approaches have achieved notable progress in their respective focused technical aspects, these methods, as isolated technical modules, are incapable of establishing systematic collaborative scheduling and dynamic orchestration mechanisms among multi-source data, multiple categories of algorithms, and multi-level operation and maintenance tasks. As a result, the overall effectiveness of intelligent bridge operation and maintenance is constrained by a structural contradiction characterized by strong individual modules but a weak system-level integration. This contradiction leads to an excessive reliance on manual interfacing in the operation of bridge O&M systems, and significant bottlenecks persist in areas such as multi-modal fusion, task generalization, deep utilization of domain knowledge, and interpretable decision-making. There is thus an urgent need for a methodological framework that can formalize executable actions, physical mechanisms, expert evidence, and operational context into a composable and traceable reasoning process.

In recent years, the rapid development of large language models has offered a novel technical pathway for addressing the aforementioned challenges of multi-task system integration. Large-scale pre-trained language models represented by GPT series [13], DeepSeek series [14], and LLaMA series [15] have continuously achieved breakthroughs in dimensions such as model parameter count, context window length, and reasoning capability [16], demonstrating powerful semantic understanding, contextual reasoning, and few-shot (zero-shot) generalization abilities, making it possible for machines to move from perception to cognition [17]. OpenAI o1 [13] achieved significant breakthroughs in mathematics, programming, and scientific reasoning tasks through internalized chain-of-thought (CoT) reasoning, marking the evolution of LLMs from rapid response to deep thinking. At the same time, vision-language multi-modal large models (MLLMs) bridge the textual semantic space with the representation spaces of other modalities such as vision and time series through cross-modal alignment mechanisms, realizing a unified cross-modal understanding and generation framework. SAIL [18] proposed a unified Transformer architecture, integrating raw pixel encoding and language decoding into a single model, eliminating the need for independent visual encoders; Mono-InternVL [19] embeds visual knowledge into the pre-trained LLM through a multi-modal expert mixture structure, achieving end-to-end optimization of cross-modal information flow. The breakthrough in reasoning capability of LLMs has given rise to the agent paradigm centered on it. LLM agents refer to interactive systems capable of perceiving visual stimuli, language input, and other environmental basic data, and capable of producing meaningful embodied actions [20]. By integrating peripheral components such as memory modules, knowledge bases, tool invocation, and planning reasoning, the models are enabled to possess the abilities of perceiving the environment, decomposing tasks, scheduling tools, reflecting and correcting errors, and continuous learning [21]. With the increasing maturity of LLM agent technology, many LLM-based agent application studies have begun to emerge in the civil engineering field. Yan et al. [22] proposed a general-purpose AI agent framework for intelligent buildings based on the ReAct strategy, achieving 91% task completion in simulated virtual building environments and successfully deploying it in physical building control systems. Deng et al. [23] introduced the BIMgent framework, which enables autonomous building model creation through graphical user interface operations. Du et al. [24] proposed Text2BIM, an LLM-based multi-agent framework capable of converting natural language instructions into three-dimensional building models. Chen et al. [25] developed a multi-agent framework for reinforced concrete structural design, decomposing complex design tasks into subtasks collaboratively handled by multiple specialized LLM agents. Furthermore, agent technologies have also demonstrated broad application prospects in areas such as adaptive building production systems combining agents with digital twins [26], multi-agent urban management [27], and multi-agent deep reinforcement learning–based decision-making for resilient community post-disaster recovery [28].

Despite the potential demonstrated by LLMs and their driven agents in general-purpose reasoning and autonomous task execution, their direct application in highly specialized domains such as bridge operation and maintenance faces fundamental cognitive translation barriers. The core reason lies in the fact that the operational judgments of senior domain experts rely heavily on tacit knowledge—that is, intuitive expertise and contextual cognition that are difficult to articulate explicitly, lacking mathematical formulations or rule-based abstractions. Existing engineering application attempts often inject domain knowledge into LLMs in a flat manner via natural language prompts, which, constrained by the model's context window and parametric memory hallucinations, fail to characterize the complex hierarchical, causal, or logical interrelationships among entities. Moreover, the non-quantifiable, experiential knowledge embedded in extensive standard specifications, inspection reports, and expert heuristics—if not systematically transformed from tacit cognition into computable structures—prevents LLM agents from simulating the experts' unstructured reasoning, leaving a persistent cognitive gap at critical decision-making junctures. Therefore, systematically adapting the general-purpose reasoning capabilities of LLMs to the knowledge constraints and task characteristics of bridge operation and maintenance necessitates, in particular, addressing the digital representation of expert experience and tacit cognition.

Knowledge graph (KG) and Retrieval-Augmented Generation (RAG) offer LLM agents external knowledge access, alleviating hallucinations and timeliness issues. Lewis et al. [29] first proposed RAG, combining parametric and non-parametric memory to dynamically access external knowledge, mitigating timeliness, explainability, and hallucination issues. Edge et al. [30] proposed Graph RAG using community summaries, surpassing vector RAG for global queries. Luo et al. [31] introduced RoG, unifying planning, retrieval, and reasoning over KG paths to ensure explainable results, demonstrating deep KG-LLM synergy. In bridge O&M, KG has been applied to knowledge management, inspection, and maintenance. Zhang et al. [32] built a Bridge Maintenance KG (BMKG) via hybrid methods, using rule- and ontology-based reasoning for automatic completion and consistency. Wang et al. [33] integrated LLMs with CoT into BMKG (BMKG-DCoT), simulating expert workflows to ensure transparent reasoning paths. Luo et al. [34] constructed a bridge inspection KG and complex QA corpus with a three-level relation method, supporting domain adaptation. Fang et al. [35] proposed a KG for bridge pile foundation construction quality management, using BiLSTM-CRF for entity/relation extraction from multi-source texts with over 85% accuracy. Gao et al. [36] explored GraphSAGE for embedding learning of bridge maintenance KG, enabling link prediction and node classification. For slab bridge inspection, a hierarchical KG [37] organizes knowledge at granular levels, showing specific structure-type value. For concrete crack detection, a KG-based bibliometric analysis [38] revealed hotspots, collaboration, and evolution trends. However, even when expert knowledge is structured through knowledge graphs and RAG techniques and subsequently injected into LLM-based agents, the reasoning process still lacks an interpretable mapping to the underlying physical mechanisms. The internal decision-making mechanisms of current LLM- or agent-based approaches remain, in essence, driven by statistical associations. Regardless of how rich the domain facts provided by the knowledge graph may be, the model's inferential pathways are still built upon probabilistic fittings of linguistic patterns, and thus fail to genuinely engage with the fundamental physical laws governing structural mechanics, material constitutive relations, and load–response behaviors. This entails that, when such agents are deployed in high-stakes scenarios such as bridge safety assessment and retrofitting decision-making, the derived reasoning chains, although semantically coherent, still lack an explicable correspondence to physical and mechanical principles. The statistically driven decision-making paradigm is inherently inadequate to meet the stringent requirements for safety, traceability, and trustworthiness in high-risk engineering contexts.

To address the above challenges, this paper proposes a knowledge-enhanced autonomy formalization methodology for bridge O&M, and applies it to construct BOMMA (Bridge O&M Multi-Modal Multi-Agent System). With an LLM serving as the central cognitive engine, the methodology incorporates a mechanics-aware action-space and management mechanism, which consolidates dispersed inspection, monitoring, and assessment tools and methodologies into standardized, semantically dispatchable functional units. Through an interactive multi-agent collaborative workflow, it enables autonomous task planning, parallel scheduling, execution-with-reflection, and result synthesis driven by natural-language maintenance instructions. Furthermore, by leveraging a knowledge-graph-augmented domain knowledge module by long-range attention mechanisms, it injects both bridge O&M expert heuristics into the agent's O&M decision-making close-loop in the form of structured semantic embeddings. This design yields traceable evidence chains of expert reasoning and physically consistent, interpretable inferential pathways for root-cause analysis, maintenance decision-making, and safety evaluation.

## 2 Methodology

This study formulates the demands for intelligent bridge operation and maintenance (O&M) as the following problem. Given an integrated O&M agent framework empowered by a core large language model (LLM) that sequentially predicts token probability densities in a reasoning sequence according to prerequisites, the framework is capable of performing streaming O&M instruction understanding and decomposition, explainable O&M action scheduling for task execution, and structured expert knowledge-driven O&M decision-making. The overall problem formulation is mathematically expressed in Equation (1):

$$\begin{aligned} & p(\boldsymbol{X}_{\mathrm{Dec}}, \boldsymbol{X}_R, \boldsymbol{X}_O, \boldsymbol{X}_A | \boldsymbol{X}_{\mathrm{PM}}, \boldsymbol{X}_M, \boldsymbol{X}_I, \mathcal{G}) \\ & \approx p(\boldsymbol{X}_{\mathrm{Dec}}, \boldsymbol{X}_R, \boldsymbol{X}_O, \boldsymbol{X}_A | \boldsymbol{X}_{\mathrm{PM}}, \boldsymbol{X}_M, \boldsymbol{X}_I, \mathcal{G}_R) \\ & = p(\boldsymbol{X}_{\mathrm{Dec}} | \boldsymbol{X}_R, \boldsymbol{X}_O, \boldsymbol{X}_A, \boldsymbol{X}_{\mathrm{PM}}, \boldsymbol{X}_M, \boldsymbol{X}_I, \mathcal{G}_R) \cdot p(\boldsymbol{X}_R | \boldsymbol{X}_O, \boldsymbol{X}_A, \boldsymbol{X}_{\mathrm{PM}}, \boldsymbol{X}_M, \boldsymbol{X}_I) \\ & \cdot p(\boldsymbol{X}_O | \boldsymbol{X}_A, \boldsymbol{X}_{\mathrm{PM}}, \boldsymbol{X}_M, \boldsymbol{X}_I) \cdot p(\boldsymbol{X}_A | \boldsymbol{X}_{\mathrm{PM}}, \boldsymbol{X}_M, \boldsymbol{X}_I) \end{aligned} \tag{1}$$

where $p(\bullet)$ denotes the conditional probability distribution predicted by an LLM with its learnable parameter $\boldsymbol{\varphi}$. $\boldsymbol{X}_{\mathrm{Dec}}, \boldsymbol{X}_R, \boldsymbol{X}_O$ and $\boldsymbol{X}_A$ represent the token sequences of decision, action reasoning, action output, and action scheduling reasoned by the LLM, respectively. $\boldsymbol{X}_M, \boldsymbol{X}_{\mathrm{PM}}, \mathcal{G}$ respectively refer to the token sequences of the agent's inherent memory, physical-mechanical action space, and structured expert knowledge base. $\boldsymbol{X}_I$ is the token sequence of an O&M instruction from users. Note that the first equality holds approximately because the generation of O&M decision $\boldsymbol{X}_{\mathrm{Dec}}$ from all the preceding reasoning $(\boldsymbol{X}_R, \boldsymbol{X}_O, \boldsymbol{X}_A)$ only needs to be conditional on the related knowledge $\mathcal{G}_R$ rather than the whole base $\mathcal{G}$, and it is only conditional to the final decision-making stage.

To resolve the aforementioned demands, this study proposed a knowledge-enhanced autonomy formalization methodology for bridge O&M, and applies it to construct a multi-modal, multi-agent system for bridge O&M—termed BOMMA (Bridge O&M Multi-Modal Multi-Agent System). The main idea of the methodology, illustrated in Figure 1, is responding to O&M instructions with an "comprehension, decomposition, execution, decision" close-loop, which leverages memory, physical-mechanical action space, expert knowledge base, and interactive agentic workflow, as a solution for Equation (1). These components fundamentally distinguish the agent from the working paradigm of general LLMs—which operate in a passive response mode, engage in text-based question answering, lack explicit constraints, remain unbiased by prior context, and start each new session with zero information. In contrast, the proposed autonomy is enabled to interpret user intent with memory at the outset of new O&M assignments, initiating O&M task planning and deployment execution grounded in physical-mechanical information, and formulate O&M decision assimilating evidence chains. Notably, the action space and knowledge base of BOMMA are constructed automatically, relying on the agentic system's core LLM as the foundational engine.

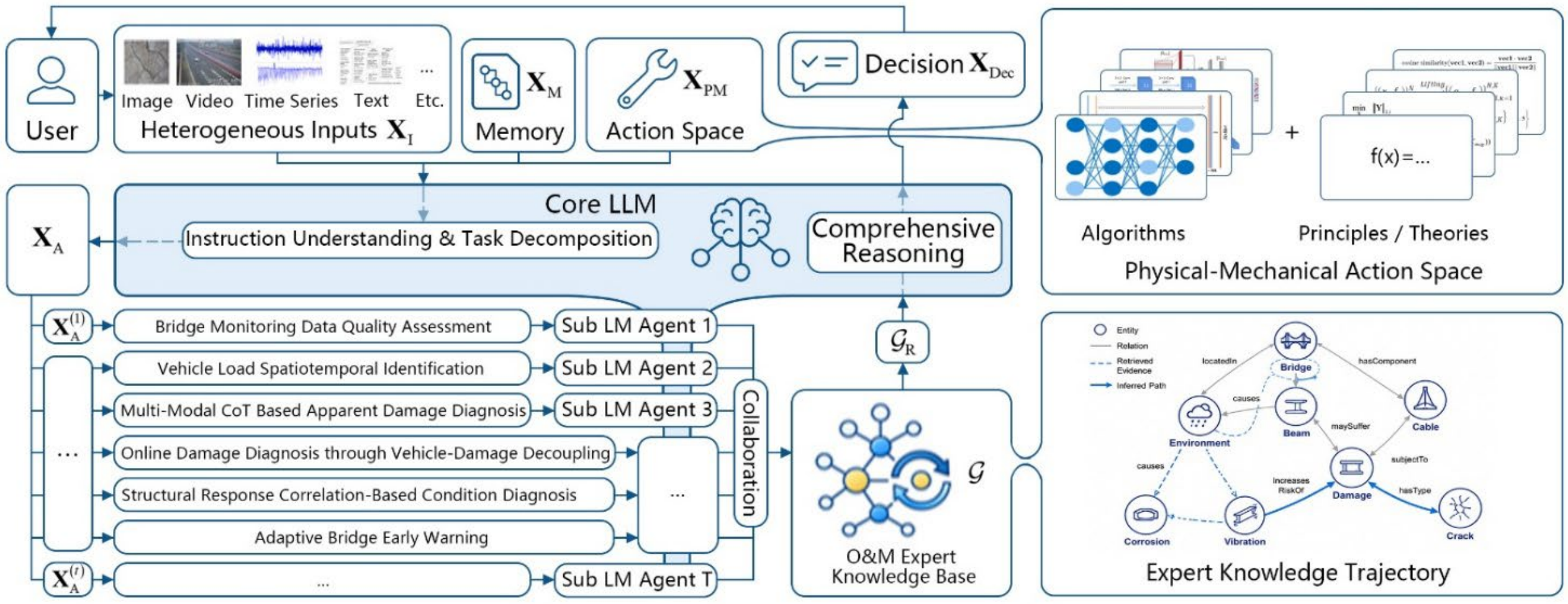

Figure 1 Workflow of Bridge O&M Multi-Agent System.

The core contribution and novelty of BOMMA resides in an autonomy formalization mechanism for bridge O&M with a three complementary layers rather than a collection of isolated methods and modules. At the organizational level, a highly autonomous multi-agent framework endows the system with self-directed management, scalable expansion, deliberative task planning, and adaptive serial–parallel allocation of the action space, enabling robust handling of composite O&M scenarios. Such autonomy is grounded in the execution level of a physical–mechanical enhanced action space, which supplies rigorous and interpretable reasoning chains for action understanding, scheduling, and outcome analysis across the full task spectrum—from signal preprocessing and damage identification to condition assessment, and maintenance decision-making—thereby anchoring agent behaviors to consistent physical principles rather than opaque heuristics. Complementing the above, the experiential knowledge level organizes, through structured semantic embeddings, the tacit and non-quantifiable expertise accumulated in engineering practice, rendering it as traceable evidence chains that underpin downstream decisions such as fault-cause analysis and maintenance recommendation. Together, the autonomy provides organizational capacity for O&M disposal, enabling BOMMA to autonomously determine its behaviors that fulfill the O&M objectives.

## 2.1 Physical-Mechanical Action Space for O&M Action Comprehension

General-purpose LLMs and their derivative agent-based approaches do not emphasize theoretical method understanding or explicit action-trajectory tracing. Consequently, such paradigms lack the reliability and interpretive guidance required for bridge O&M—a domain that places significant emphasis on logical accuracy and decision safety—and are therefore unsuitable for practical engineering deployment. Therefore, this study integrates physical-mechanical mechanisms into the agent's action space, in order to provide more rigorous and interpretable reasoning for action understanding, scheduling, and outcome analysis in bridge O&M tasks. Note that the creation of BOMMA's action space is automatically directed by its core LLM, which is illustrated in Appendix A in detail.

In the beginning of the close-loop, an O&M instruction from users is comprehended and decomposed into a subtask sequence according to the physical-mechanical action space, shown in Equation (2):

$$\begin{aligned} p(\boldsymbol{X}_A|\boldsymbol{X}_{\text{PM}},\boldsymbol{X}_M,\boldsymbol{X}_I) &= p\left(\bigcup_{t=1}^{T}\boldsymbol{X}_A^{(t)}\middle|\bigcup_{t=1}^{T}\boldsymbol{X}_{\text{PM}}^{(t)},\boldsymbol{X}_M,\boldsymbol{X}_I\right) \\ &= \prod_{t=1}^{T} p\left(\boldsymbol{X}_A^{(t)}\middle|\boldsymbol{X}_A^{(1)},\dots,\boldsymbol{X}_A^{(t-1)},\boldsymbol{X}_{\text{PM}}^{(t)},\boldsymbol{X}_M,\boldsymbol{X}_I\right) \end{aligned} \tag{2}$$

where $\boldsymbol{X}_A$ equalizes to a series of independent action token sequences $\left\{\boldsymbol{X}_A^{(1)},\dots,\boldsymbol{X}_A^{(T)}\right\}$ for the accomplishment of an O&M instruction, whose corresponding physical-mechanical principles $\left\{\boldsymbol{X}_{\text{PM}}^{(1)},\dots,\boldsymbol{X}_{\text{PM}}^{(T)}\right\}$ are considered available for solving the proposed instruction. The agent's memory $\boldsymbol{X}_M$ assists the instruction comprehension with restrictive and supplementary information, whose content is detailed in Table B.2. Note that each action is scheduled conditional on its corresponding part of physical-mechanical principles and the preceding scheduled actions. The agent then obtains action output from each action based on the task decomposition, illustrated in Equation (3):

$$\begin{aligned} p(\boldsymbol{X}_O|\boldsymbol{X}_A,\boldsymbol{X}_{\text{PM}},\boldsymbol{X}_M,\boldsymbol{X}_I) &= p\left(\bigcup_{t=1}^{T}\boldsymbol{X}_O^{(t)}\middle|\bigcup_{t=1}^{T}\boldsymbol{X}_A^{(t)},\bigcup_{t=1}^{T}\boldsymbol{X}_{\text{PM}}^{(t)},\boldsymbol{X}_M,\boldsymbol{X}_I\right) \\ &= \prod_{t=1}^{T} p\left(\boldsymbol{X}_O^{(t)}\middle|\boldsymbol{X}_A^{(t)},\boldsymbol{X}_{\text{PM}}^{(t)},\boldsymbol{X}_M,\boldsymbol{X}_I\right) \end{aligned} \tag{3}$$

where $\boldsymbol{X}_O^{(t)}$ denotes the action output from the *t*-th subtask, which is only conditional on its assigned

$\boldsymbol{X}_A^{(t)}$ and the corresponding $\boldsymbol{X}_{\mathrm{PM}}^{(t)}$ without any other portion of the action space.

This integration of physical-mechanical principles fundamentally relies on long-range reasoning capability—a domain in which LLM agents, by virtue of the self-attention mechanism inherent to the Transformer architecture, substantially mitigate the information loss existing in previous natural language processing (NLP) methodologies that plague long-sequence token prediction during action reasoning. The action reasoning is illustrated in Equation (4):

$$\begin{aligned} p(\boldsymbol{X}_R|\boldsymbol{X}_O,\boldsymbol{X}_A,\boldsymbol{X}_{\mathrm{PM}},\boldsymbol{X}_M,\boldsymbol{X}_I) &= p\left(\cup_{t=1}^{T}\boldsymbol{X}_R^{(t)}\middle|\cup_{t=1}^{T}\boldsymbol{X}_O^{(t)},\cup_{t=1}^{T}\boldsymbol{X}_A^{(t)},\cup_{t=1}^{T}\boldsymbol{X}_{\mathrm{PM}}^{(t)},\boldsymbol{X}_M,\boldsymbol{X}_I\right) \\ &= \prod_{t=1}^{T}p\left(\boldsymbol{X}_R^{(t)}\middle|\boldsymbol{X}_O^{(t)},\boldsymbol{X}_A^{(t)},\boldsymbol{X}_{\mathrm{PM}}^{(t)},\boldsymbol{X}_M,\boldsymbol{X}_I\right) \end{aligned} \tag{4}$$

where $\boldsymbol{X}_R^{(t)}$ denotes the action reasoning of the $t$-th subtask. Note that $\boldsymbol{X}_R^{(t)}$ is only correlating to its own subtask output, its assigned action and the corresponding physical-mechanical principles. When LLMs compute the predictive probability of the next token in an expanding context sequence of an O&M action based on physical-mechanical principles, they calculate attention across all preceding tokens, particularly including those from physical-mechanical principles, and performs a weighted sum according to the value matrix [39]. This paradigm ensures that the information propagation path length between any two tokens is uniform, irrespective of their absolute positional distance in the sequence. Moreover, it circumvents the information compression inevitably incurred when representing arbitrarily long sequence histories with fixed-dimensional implicit parameters—a limitation that plagued earlier NLP models and created a transmission bottleneck where information attenuated with increasing distance, therefore maintaining the guidance of physical-mechanical principles in long-term token generation.

Following the methodology described above, BOMMA's action space primarily comprises the physical-mechanical actions listed in Table 3, and the details of these physical-mechanical actions are discussed in the following:

**Bridge Monitoring Data Quality Assessment [40].** The action of bridge monitoring data quality assessment takes structural monitoring data $z = \{z^{(1)}, \dots, z^{(n_z)}\}$ as input and produces a quality evaluation score $QES$ for the entire time series. An outlier-resistant conditional diffusion model is trained as an implicit univariate autoregressive predictor to learn the conditional distribution of each observed data point conditioned on its temporal context. Subsequently, each data point is assigned an outlier probability that accounts for both predictive uncertainty $\sigma_p^{(t)}$ and the data noise $\sigma_e$, thereby quantifying its degree of "outlier-ness":

$$p_o(z^{(t)} \mid \hat{z}^{(t)}, \sigma_p^{(t)}, \sigma_e) = \left[1 - 2\Phi\left(-\left|\frac{z^{(t)} - \hat{z}^{(t)}}{\sqrt{\sigma_p^{(t)} + \sigma_e}}\right|\right)\right]^{n_w} \tag{5}$$

where $\Phi(\cdot)$ denotes the cumulative distribution function of the standard normal distribution. This probability indicates the likelihood that the absolute residuals of all $n_w$ data points are smaller than the current $|z^{(t)} - \hat{z}^{(t)}|$. Observations satisfying $p_o > 0.5$ are classified as low-quality data and replaced with their predictive means to obtain the cleaned series. Finally, given $n_q$ identified low-quality data among the $n_e$ evaluated data points, the quality evaluation score *QES* for global data quality assessment is computed as:

$$\mathrm{QES} = 1 - \frac{2 \times \frac{n_q}{\kappa n_e} \times \frac{1}{n_q} \sum_{i=1}^{n_q} p_o(z^{(i)})}{\frac{n_q}{\kappa n_e} + \frac{1}{n_q} \sum_{i=1}^{n_q} p_o(z^{(i)})} \tag{6}$$

where $\kappa$ is a dimensionless scaling factor and $n_c$ denotes the context-window length. This metric aggregates the scaled fraction of low-quality data and their mean outlier probability via a harmonic-mean-style pooling operation.

**Vehicle Load Spatiotemporal Distribution Identification**[4]**.** This action takes multi-camera bridge surveillance videos with the weighbridge detected load data as input, and outputs whole field vehicle load spatiotemporal distribution. It proceeds in three stages: vehicle exterior identification, cross-camera tracking and load associating. Firstly, a YOLO [42] detector is run on the video keyframes, and the boxes whose detection confidence exceeds a preset threshold are kept as detected vehicles. Then a HardNet-based tracker performs cross-camera vehicle appearance matching and trace tracking, which is trained by the loss function illustrated in Equation (7):

$$
\begin{cases}
\mathcal{L}_{\mathrm{HN}} = \frac{1}{n_v}\sum_{i=1}^{n_v} \max\left(0,1 + d(\boldsymbol{a}_i, \boldsymbol{p}_i) - \min\left(d\left(\boldsymbol{a}_i, \boldsymbol{p}_{j_{\min}}\right)\right), d\left(\boldsymbol{a}_{k_{\min}}, \boldsymbol{p}_i\right)\right) \\
j_{\min} = \arg\min_{j=1,\ldots,n_v, j\neq i} d\left(\boldsymbol{a}_i, \boldsymbol{p}_j\right) \\
k_{\min} = \arg\min_{k=1,\ldots,n_v, k\neq i} d(\boldsymbol{a}_k, \boldsymbol{p}_i)
\end{cases}
\tag{7}
$$

where $n_v$ denotes the number of vehicles, $\boldsymbol{a}_i, \boldsymbol{p}_i$ respectively denote a vehicle's feature point descriptor and its corresponding mapped descriptor. $\boldsymbol{p}_{j_{\min}}$ denotes the nearest non-feature descriptor to $\boldsymbol{a}_i$, and $\boldsymbol{a}_{k_{\min}}$ denotes the one nearest to $\boldsymbol{p}_i$. $d(\boldsymbol{a}_k, \boldsymbol{p}_i)$ is the distance metric. After which the vehicles are traceable by exteriors in the whole field, their loads are associated with their traces by the instants of weighbridge crossing, realizing whole field vehicle load spatiotemporal distribution.

**Multi-Modal Chain of Thought Based Apparent Damage Diagnosis [6].** The action of apparent damage identification takes structural damage images and user query texts as input, through image encoding, cross modal feature alignment, and chain of thought reasoning, finally outputs damage categories and feature descriptions. The model adopts the ViT-G module as the image encoder and uses the Q Former module to accomplish the image-text cross-modal semantic alignment task. This module introduces a set of learnable query vectors, which select information highly relevant to text semantics from image features via a cross attention mechanism, and Equation (8) shows the computation process:

$$
u^{(i)} = \mathrm{Attn}(U_q^{(i)}, U_k^{(i)}, U_v^{(i)}) = \mathrm{softmax}\left(\frac{U_q^{(i)} U_k^{(i)\top}}{\sqrt{n_a}}\right) U_v^{(i)} \tag{8}
$$

where $U_q^{(i)}$ is obtained by linear transformation of the learnable query vectors, $U_k^{(i)}$ and $U_v^{(i)}$ are obtained by linear transformation of image feature encodings, and the transformation matrices are learnable weights. This design enables the model to dynamically focus on damage-related regions and map image embeddings into a space semantically consistent with text embeddings, thereby achieving unified representation of multi-modal information.

To achieve logically rigorous damage description, the model introduces a text-visual fused chain-of-thought mechanism. The model provides the generated semantic segmentation map of damage as prior knowledge to the model, forming a visual reasoning process, while at the same time guiding the model via text prompts to first determine the damage category and then describe the damage characteristics. Equation (9) shows the multi-modal chain-of-thought reasoning process.

$$p(\boldsymbol{X}_{\text{out}}, \boldsymbol{X}_{\text{cot}} \mid \boldsymbol{X}_1, \boldsymbol{X}_2, \boldsymbol{X}_p, \boldsymbol{X}_s) = p(\boldsymbol{X}_{\text{out}} \mid \boldsymbol{X}_p, \boldsymbol{X}_1, \boldsymbol{X}_2, \boldsymbol{X}_s, \boldsymbol{X}_{\text{cot}}) \cdot p(\boldsymbol{X}_{\text{cot}} \mid \boldsymbol{X}_1, \boldsymbol{X}_2, \boldsymbol{X}_p, \boldsymbol{X}_s)$$
$$= \prod_{j=1}^{|X_{\text{out}}|} p_{\text{LM}}(x_{\text{out}}^{(j)} \mid \cdots) \cdot \prod_{i=1}^{|X_s|} p_{\text{LM}}(x_{\text{cot}}^{(i)} \mid \cdots) \tag{9}$$

where $\boldsymbol{X}_1$ and $\boldsymbol{X}_2$ are text and image inputs respectively, $\boldsymbol{X}_p$ corresponds to the chain-of-thought prompt, $\boldsymbol{X}_s$ denotes the semantic segmentation map, $\boldsymbol{X}_{\text{cot}}$ is the text reasoning content, and $\boldsymbol{X}_{\text{out}}$ is the model output. By decomposing the problem into intermediate steps through multi-modal chain-of-thought, more prior knowledge is provided for the final reasoning result, improving the accuracy of apparent damage identification.

**Online Damage Diagnosis through Vehicle-Damage Decoupling [7].** The action takes multi-type sensor data and lane surveillance video data as inputs, and outputs the overall health condition of the bridge. The action trains an encoder-decoder architecture model named GAS-GRU, which uses gate recurrent unit (GRU) as its backbone. The model treats external vehicle loads as the physical representation corresponding to the latent space, learning the spatial-temporal distribution of vehicle information on the bridge. Considering the differences in sensor data types and locations, an attention mechanism is introduced to allocate weights, enabling better inference of the spatial-temporal distribution of vehicles in the latent space. Subsequently, intelligent diagnosis of bridge damage is performed by calculating the residuals between the multitype response data reconstructed by the decoder and the actual monitored data, illustrated as following:

$$\begin{cases} \widehat{W}^{(t)} = (g_W(z^{(t)}), h^{(t-1)}) \\ \hat{z}^{(t)} = (g_z(\widehat{W}^{(t)}), h^{(t-1)}) \\ h^{(t)} = g_h(z^{(t)}, \widehat{W}^{(t)}, h^{(t-1)}) \end{cases} \tag{10}$$

where $\widehat{W}^{(t)}$ denotes the inferred external vehicle load matrix, and $z^{(t)}$ represents the current sensor monitoring data. $g_x(\cdot)$ and $g_w(\cdot)$ are projection functions mapping the inputs to the hidden-layer dimension $n_h$. $g_h(\cdot)$ corresponds to the hidden state update function. $g_W(\cdot)$ is a projection function mapping the inputs to the external vehicle load dimension. $\hat{z}^{(t)} \in \mathbb{R}^{n_{\text{ch}}}$ denotes the reconstructed monitoring data at the current time step. $g_z(\cdot)$ is a projection function mapping the inputs to the measurement channel dimension $n_{\text{ch}}$.

Finally, the residual magnitude across the *C* sensor channels participating in the diagnosis is calculated as follows:

$$\mathcal{L}_{\text{res}}^{(c)} = \frac{1}{n_t}\sum_{t=1}^{n_t}\left\|z^{(c,t)} - \hat{z}^{(c,t)}\right\|_2^2 \quad (11)$$

where *c* denotes the sensor channel index.

Based on the training data, the action sets corresponding safety thresholds $\tau_s$ for different channels; if the residual value exceeds the safety threshold $\tau_s$, we determine that damage has occurred at the location of that sensor.

**Structural Response Correlation-Based Condition Diagnosis [41].** The action of structural condition diagnosis is based on a StarGAN-based diagnostic model for the static-dynamic response correlation state of long-span bridges. The action takes the response distributions statistically obtained from each monitoring channel as input, and is conditioned on a target response type label. Through a single generator, the model enables mutual prediction between any two types of response distributions. The generator generates a distribution of the category corresponding to the target label. The model adopts an adversarial training strategy, in which the discriminator simultaneously outputs a realism score and class probabilities, so as to force the generator to learn the intrinsic mapping relationships among the various response distributions. The training process jointly optimizes three loss terms: adversarial loss, classification loss, and reconstruction loss. The adversarial loss with gradient penalty is computed as shown in Equation (12):

$$\mathcal{L}_{\text{adv}} = \mathbb{E}_z[D_{\text{src}}(z)] - \mathbb{E}_{z,\eta}[D_{\text{src}}(\text{Gen}(z,\eta))] - \kappa_{\text{gp}}\mathbb{E}_{\hat{z}}[(\|\nabla_{\hat{z}}D_{\text{src}}(\hat{z})\|_2 - 1)^2] \quad (12)$$

where $\mathcal{L}_{\text{adv}}$ denotes the adversarial loss with gradient penalty, $z$ denotes the response distribution, $D_{src}$ denotes the realism score, $\eta$ denotes the target response label, $\text{Gen}$ is the generator, $\kappa_{\text{gp}}$ is the gradient penalty coefficient and $\hat{z}$ denotes a linear interpolation sample between the real distribution and the generated distribution.

The classification loss encourages the generator to enable the discriminator to correctly recognize the generated distribution as the target label, and its formulation is given in Equation (13):

$$\mathcal{L}_{\text{cls}} = \mathbb{E}_{z,\eta}[-\log D_{\text{cls}}(\eta \mid \text{Gen}(z,\eta))] \tag{13}$$

where $\mathcal{L}_{\text{cls}}$ denotes the classification loss of the generated sample. The reconstruction loss, through cycle-consistency constraints, ensures that the generation process preserves the main content information of the distribution.

$$\mathcal{L}_{\text{rec}} = \mathbb{E}_{z,\eta,\eta'}[\|z - \text{Gen}(\text{Gen}(z,\eta),\eta')\|_1] \tag{14}$$

where $\mathcal{L}_{\text{rec}}$ denotes the reconstruction loss enforcing cycle consistency and $\eta'$ is the original label of the input distribution $z$.

After training, a response distribution $z$ of a certain type is fed into the generator with a specified target label $\eta$, yielding a prediction $\hat{z} = \text{Gen}(z,\eta)$ of the target response distribution. Subsequently, the Earth Mover's Distance (EMD) between the predicted distribution and the real distribution is computed, and the time series of this distance is tracked on a daily basis. When the EMD sequence corresponding to a particular sensor exhibits a persistent and significant deviation from the baseline level, it indicates that the correlation pattern among the responses at that location has undergone an abrupt change, thereby leading to the determination that the structural state at that location has changed, and thus achieving damage detection and localization.

**Adaptive Bridge Early Warning.** The action of adaptive early warning of bridge failure risk takes structural health monitoring data $z^{(1:t)}$ and the action history $a^{(1:t-1)}$ as input and produces an early warning decision $a^{(t)}$ at each decision epoch $t$. Since the latent structural state $h^{(t)}$ cannot be directly observed, a belief state $b^{(t)}$ is defined to represent the posterior distribution of the structural state conditioned on all monitoring data and actions so far:

$$b^{(t)} = p(h^{(t)} \mid z^{(1:t)}, a^{(1:t-1)}) \tag{15}$$

By propagating this belief state through a probabilistic state transition model, the failure risk $\rho^{(t)}$ can be quantified as the cumulative failure probability over a prescribed horizon $n_s$:

$$\begin{cases} \rho^{(t)} = 1 - \prod_{s=1}^{n_s}\left(1 - p_f^{(t+s)}\right) \\ p_f^{(t+s)} = \int_{\Omega_F} p\left(h^{(t+s)} \mid b^{(t)}, a^{(t:t+s-1)}\right) dh^{(t+s)} \end{cases} \tag{16}$$

where $p_f^{(t+s)}$ denotes the conditional failure probability during the $s$-th future decision interval; $\Omega_F$ denotes the failure domain. The early warning decision is triggered through a threshold-based strategy, and the threshold $\tau_w$ is optimized through pre-posterior analysis over the prospective structural observations predicted by a Bayesian dynamic linear model:

$$\tau_w^{*(t)} = \operatorname{argmin}_{\tau_w} \mathbb{E}_{\hat{z}^{(t+1:t+n_s)}}\left[\sum_{s=1}^{n_s} \gamma^s J(h^{(t+s)}, a^{(t+s)}(\tau_w)) \mid b^{(t)}\right] \tag{17}$$

where $n_s$ is the remaining service horizon; $\hat{z}^{(t+1:t+n_s)}$ denotes the prospective structural observations predicted by a Bayesian dynamic linear model; $J(h^{(t+s)}, a^{(t+s)})$ denotes the cost model that quantifies the economic consequences of each action, including reconstruction, life loss, traffic interruption, and time loss costs. $\gamma$ is the discount factor; and $a^{(t+s)}$ is the threshold-based warning action induced by the risk measure. At each decision epoch, the bridge manager collects the monitoring data within the current interval and updates the belief state via Bayes' theorem, and the warning threshold is then optimized based on the updated belief and the remaining horizon. An early warning is issued and the closure measure is implemented when the predicted failure risk $\rho^{(t)}$ exceeds the optimized threshold $\tau_w^{*(t)}$; otherwise, the bridge keeps open and the process repeats upon the arrival of new monitoring data. This process forms a closed loop of updating, prediction, optimization, and decision execution, in which the warning threshold is adaptively updated to dynamically balance the structural failure risk against the warning response cost as the uncertainty and cost evolves.

**O&M Expert Knowledge Base for maintenance decision-making.** The action of the bridge O&M expert knowledge base provides expert experience and reasoning for maintenance decision-making, grounded in the bridge O&M knowledge graph system, as detailed in Section 2.2.

Table 3 Major Content of BOMMA's Physical-Mechanical Action Space

| Primary Function | Abbreviation | Action |
|---|---|---|
| Data Pre-Processing | DQA | Probabilistic bridge monitoring data quality assessment |
| Load identification | VEH | Vehicle Load Spatiotemporal Distribution Identification |
| Damage Diagnosis | VIS | Multi-Modal CoT Based Apparent Damage Diagnosis |
| | VDD | Online Damage Diagnosis through Vehicle-Damage Decoupling |
| | STG | Structural Response Correlation-Based Condition Diagnosis |
| Decision Making | AEW | Adaptive Bridge Early Warning |
| | KMD | O&M Expert Knowledge Base for maintenance decision-making |
| Others | \ | Other essential actions for bridge O&M, such as action creator (ACR), weather checking (WTH), web searching (WEB), etc. |

## 2.2 Structured O&M Knowledge for Expert Decision Trajectory

This study transforms practical engineering knowledge into structuralize semantic embeddings, providing traceable expert thinking evidence chains for decision-making. The bridge O&M knowledge system contains a certain volume of experiential, tacit, non-quantifiable expert knowledge, and these contents are fragmented and scattered within massive historical data. Ordinary retrieval or enhanced retrieval can only locate scattered relevant words or passages when facing complex problems, unable to directly form complete reasoning chains. Therefore, it is necessary to transform knowledge into explicitly structured knowledge graphs, mining the reasoning logic threads implied within expert experience and thinking.

LLM-based agents, by virtue of their few-shot and zero-shot learning capabilities [43], are naturally adapted to extracting structured knowledge graph tuples from unstructured data. The implementation mechanism of this capability is mainly rooted in the generalized semantic understanding and reasoning ability of LLMs, which is gained by pre-training at scale on massive text corpora spanning diverse domains of human knowledge and behaviors—encompassing a wide range of intents, processes, and outcomes. This broad data exposure and fitting can only be achieved by LLMs containing sufficient learnable parameters [44], enabling them not only to learn broad knowledge, but also to capture task-independent logical relationships and language patterns in context organization. Utilizing this highly transferable implicit representation, LLMs are able to extract tuples from context on a passage-by-passage basis relying solely on prompts, and the extracted tuples inherently form a graph structure, enabling fully automated knowledge graph construction.

Leveraging the aforementioned few-shot and zero-shot learning capabilities, this study establishes the BOMMA knowledge module of automatically constructing knowledge graphs $\mathcal{G}$ based on unstructured historical bridge inspection reports issued by highly experienced human experts, in order to build scattered knowledge into evidence chains from identifications to assessments. The auto-construction is formalized in Equation (18):

$$\begin{cases} p(\mathcal{G}|\boldsymbol{X}_{\mathrm{Pr}}, \boldsymbol{X}_q) = p\left(\bigcup_{k=1}^{K} \boldsymbol{X}_{vrv}^{(k)} \,|\boldsymbol{X}_{\mathrm{Pr}}, \bigcup_{k=1}^{K} \boldsymbol{X}_q^{(k)}\right) = \prod_{k=1}^{K} p\left(\boldsymbol{X}_{vrv}^{(k)} \middle| \boldsymbol{X}_{\mathrm{Pr}}, \boldsymbol{X}_q^{(k)}\right) \\ \boldsymbol{X}_{vrv}^{(k)} = \left\{v_s^{(k)}, r^{(k)}, v_o^{(k)}\right\} \end{cases} \tag{18}$$

where $\boldsymbol{X}_{vrv}^{(k)}$ represents the token sequence of the $k$-th group of graph tuple, in which $v_s^{(k)}, r^{(k)}, v_o^{(k)}$ respectively represent the tokens of its subject node, relation, object node, extracted from the token sequence of the $k$-th passage $\boldsymbol{X}_q^{(k)}$ of the entire sequence of unstructured O&M text data $\boldsymbol{X}_q$ guided by the prompt token sequence $\boldsymbol{X}_{\mathrm{Pr}}$. Note that the prompt urges the nodes to be categorized into a four-layer coupled O&M domain ontology: structural entity layer (defining bridge components such as girders, piers, etc.), defect/damage evolution layer (categorizing defect types such as cracks, corrosion, etc.), hazard constraint layer (describing environmental factors that trigger defects such as landslides, floods), and O&M disposal layer (specifying targeted repair and maintenance measures). Every node and relation contains its timestamp as a key attribute for the tracing of evolution.

During the decision stage, the agent matches nodes (entities or communities) by their semantic embedding relevance to the gathered action reasoning, which contains O&M decision trajectories from directly related starting nodes to historical disposal, illustrated in Equation (19):

$$\begin{cases} \mathcal{V}_I = \left\{v^{(i)} \middle| s\left(\boldsymbol{X}_R, v^{(i)}\right) \geq \lambda\right\} \\ s\left(\boldsymbol{X}_R, v^{(i)}\right) = Emb^{\top}(\boldsymbol{X}_R)\,\mathrm{Emb}\left(v^{(i)}\right) \\ v^{(i)} \in \left\{v_s^{(1)}, \dots, v_s^{(K)}, v_o^{(1)}, \dots, v_o^{(K)}\right\} \end{cases} \tag{19}$$

where $\mathcal{V}_I$ denotes a set of starting knowledge nodes $v^{(i)}$ whose relevance to the action reasoning $\boldsymbol{X}_R$ exceed the threshold $\lambda$, $s(\cdot)$ represents the relevance metric for a pair of embeddings. Further indexing of the downstream layer O&M knowledge connected by graph structure is performed based on $\mathcal{V}_I$, obtaining a subgraph containing expert knowledge trajectories related to the O&M assignment, which is illustrated in Equation (20):

$$\mathcal{G}_R = \{\mathcal{V}_R, \mathcal{R}_R\} = \cup_{l=0}^{L} \mathcal{N}^{(l)}(\mathcal{V}_I) \tag{20}$$

where $\mathcal{G}_R$ denotes the retrieved O&M knowledge subgraph (a union of L-hop connection expansions based on $\mathcal{V}_I$, containing all the nodes $\mathcal{V}_R$ related to $\boldsymbol{X}_R$ and their relationship net $\mathcal{R}_R$), $\mathcal{N}^{(l)}(\cdot)$ represents the L-hop connection expansion retrieval. The agent reasons the final decision based on this O&M knowledge subgraph and all the preceding reasoning, which completes the workflow proposed in Equation (1).

This mechanism is aligned with bridge O&M in the following key respects: Within the standards, inspection reports, and various other document types in the bridge O&M domain, a substantial body of experiential expert knowledge exists that cannot be readily quantified or analyzed through explicit or numerical methods. LLM agents can structurally manage such knowledge via a knowledge graph approach. By leveraging the information propagation network inherent to the graph structure and the natural language embedding capabilities of LLMs, these agents can systematically explore and extract decision-relevant evidence, effectively parsing the reasoning processes underlying bridge O&M expert thinking. This, in turn, provides traceable knowledge bases and discriminable reasoning traces to support expert decision-making—including cause analysis and maintenance recommendations.

### 2.3 Interactive Agentic Workflow for Stabilized Close-Loop O&M Responding

Standalone LLMs often struggle to escape the black-box generation paradigm, resulting in limited controllability over reasoning trajectories. Moreover, in the absence of robust factual grounding, they are particularly prone to hallucination in professional domains such as bridge O&M. Building upon an interactive agentic workflow, a multi-agent framework with strong autonomy and high interpretability is established in this study to alleviate the above problem, which comprehensively integrates the principles of instruction following [45], in-context learning, and action space understanding and scheduling of LLMs.

The core enhancement introduced by the interactive agentic workflow lies in augmenting LLMs with knowledge modules, memory modules, and physical-mechanical action spaces, thereby transforming them from passive responders into autonomous actors. Through the intervention hooks of these modules, composite O&M tasks are decomposed into several explicit stages. Factual constraints—including memory $\boldsymbol{X}_M$ and physical-mechanical action space $\boldsymbol{X}_{\mathrm{PM}}$, action output $\boldsymbol{X}_O$, expert knowledge $\mathcal{G}_R$—are respectively injected at the reasoning stage of task decomposition, O&M execution, decision-making, providing automatic intermediate interventions that inform the following stages of reasoning. This design shortens the span of each individual black-box reasoning step and reduces its inherent complexity. Explainability therefore arises from the deliberate placement of knowledge constraints within the reasoning process, rather than from a post hoc explanation of an unconstrained response.

Pseudocode 1 Workflow of BOMMA from Instruction to Decision

**Input:** Token sequence of O&M instruction $\boldsymbol{X}_I$.

**Output:** Token sequences of O&M decision $\boldsymbol{X}_{\mathrm{Dec}}$, action reasoning $\boldsymbol{X}_R$, action output $\boldsymbol{X}_O$ and O&M action scheduling $\boldsymbol{X}_A$.

**Requirement:** Token sequences of memory $\boldsymbol{X}_M$, physical-mechanical action space $\boldsymbol{X}_{\mathrm{PM}}$, and structured expert knowledge $\mathcal{G}$.

# Step 1 O&M Task Decomposition

Perform task decomposition and obtain $\boldsymbol{X}_A$ as Equation (2);

# Step 2 O&M Execution

**While** not meeting the reflection termination conditions*:

    Assign $\boldsymbol{X}_A$ to T subagents and obtain the action output $\boldsymbol{X}_O$ as Equation (3);

    Subagents receive $\boldsymbol{X}_O$ and obtain action reasoning $\boldsymbol{X}_R$ as Equation (4);

**End While**

# Step 3 O&M Decision Making

**If** Decision-making is argued in $\boldsymbol{X}_I$:

Retrieve expert knowledge $\mathcal{G}_R$ as Equation (19)-(20);

Perform O&M decision making and obtain $\boldsymbol{X}_{\text{Dec}}$ as Equation (1);

**End if**

**Return** $\boldsymbol{X}_{\text{Dec}}$, $\boldsymbol{X}_R$, $\boldsymbol{X}_O$ and $\boldsymbol{X}_A$.

*The reflection termination conditions of subagents are illustrated in Appendix D.

Based on this mechanism, BOMMA establishes a workflow illustrated in Pseudocode 1. This workflow not only enhances the explainability and visibility of the overall task process, but also enables autonomous revision of subsequent steps based on verification feedback. It thereby establishes a closed-loop interaction cycle—comprising understanding, scheduling, reflection, and decision-making—that effectively suppresses the hallucination tendencies of general LLMs while simultaneously achieving autonomous execution of composite O&M tasks, i.e., making a marked improvement in the reliability of bridge O&M decisions. Moreover, the subagent-based multi-agent collaboration approach offers distinct advantages for parallelizable sub-tasks, as well as for operations that are inherently long-running—such as persistent actions common in bridge O&M (e.g., continuous sensor data acquisition over a defined period) and computationally intensive actions (e.g., video stream processing by the VEH action). By delegating these tasks to subagents, this approach avoids the efficiency bottleneck that would arise from prolonged occupation of the main agent.

## 3 Experiment

### 3.1 Implementation

The performance of BOMMA is evaluated using a total of 3,492 multi-type, multi-modal bridge O&M test instructions. These instructions comprise 2,460 simple instructions that invoke only a single bridge O&M action at a time, and 1,032 composite instructions that each schedules multiple O&M actions (2 to 5 actions per instruction), with mixed serial-parallel execution requirements. The test instructions are drawn from four distinct user sources, reflecting varying levels of familiarity with the system: (i) individuals with an in-depth understanding of the physical-mechanical mechanisms and algorithmic models underlying bridge O&M actions (Source I); (ii) individuals who do not possess full comprehension but have read the abstract of each action, mainly instructing with O&M objectives (Source II); (iii) individuals with no prior understanding, yet who are aware of the action names (Source III); and (iv) individuals who have had extensive interaction with BOMMA and have established agreed-upon user habits and minimalist instruction protocols (Source IV). The instruction set covers a broad spectrum of practical engineering scenarios, including: direct invocation of rigorous action names; calls originating from the level of physical-mechanical mechanisms or algorithmic models; calls based on action functionality; abbreviated calls; pre-agreed composite instructions; and instructions that do not explicitly mention any skill, thereby requiring the agent to make autonomous decisions. These test instructions are employed to assess the agent's action scheduling accuracy, composite instruction reasoning accuracy, user intervention frequency, and other relevant capabilities. Representative examples of these test instructions are provided in Appendix C.

Table 4 Distribution of Bridge O&M Test Instructions

| Source | Simple instruction | Composite instruction | Total |
|---|---|---|---|
| Source I | 612 | 220 | 832 |
| Source II | 928 | 200 | 1,128 |
| Source III | 920 | 200 | 1,120 |
| Source IV | 0 | 412 | 412 |
| Total | 2,460 | 1,032 | 3,492 |

BOMMA is implemented on a Linux operating system, using Node Package Manager (NPM) as the development environment and JavaScript as the programming language. On the hardware side, the server employed in this study is equipped with 2× Intel Xeon 6960P CPUs, 16×64 GB memory modules (totaling 1,024 GB), and 4× NVIDIA RTX PRO 6000 GPUs (each with 96 GB of VRAM). The algorithm models for each bridge O&M action are deployed on PyTorch, running on the aforementioned server, and are interfaced with the agent via remote application programming interfaces (APIs).

For a controlled comparison, all architectures in the following experiments adopt DeepSeek-V4.1-Flash as the core LLM, and their O&M action set, action descriptions, memory (if any) and knowledge resources, test instructions, and model settings are all the same. The compared architectures differ only in how planning, action execution, reflection, and parallelism are organized. All architecture comparisons reported in the following subsections were rerun under the same final model version and system configuration to ensure a controlled comparison.

### 3.2 Evaluation Metrics

This study primarily evaluates the performance of the proposed bridge O&M agent architecture across several key metrics, including five primary metrics that will be specifically discussed: the scheduling accuracy for each individual O&M action; the reasoning accuracy for composite instructions; the average number of user interventions required per composite instruction; the relative average execution time for composite instructions; and the physical-mechanical reasoning accuracy. For detailed performance on each bridge O&M action involved in this study, please refer to the respective references cited in Table 3.

Scheduling recall rate (Rec) for bridge O&M actions assesses the quality of automatically constructed actions. It is defined as the ratio of the number of times an action is correctly and independently scheduled by the queries to the number of independent instructions expected to invoke that action, as formalized in Equation (21):

$$\mathrm{Rec} = \frac{n_{\mathrm{TP}}}{n_{\mathrm{TP}}+n_{\mathrm{FN}}} \tag{21}$$

where $n_{\mathrm{TP}}$, $n_{\mathrm{FN}}$ respectively indicate the number of simple instructions that successfully trigger the expected actions, and those failed to do so.

Composite instruction scheduling accuracy evaluates the agent's task decomposition capability, which is further divided into scheduling precision (C-Pre), recall (C-Rec), and exact matching rate (ExactMatch). For a test composite instruction $\boldsymbol{X}_I$, C-Pre evaluates the necessity of dispatched actions, C-Rec evaluates their sufficiency, illustrated in Equation (22):

$$\begin{cases} \text{C-Pre}(\boldsymbol{X}_I) = \frac{1}{|\boldsymbol{X}_A|} \sum_{\boldsymbol{X}_A^{(i)} \in \boldsymbol{X}_A} \max_{\boldsymbol{X}_{A(\mathrm{GT})}^{(j)} \in \boldsymbol{X}_{A(\mathrm{GT})}} \boldsymbol{X}_A^{(i)} \cdot \boldsymbol{X}_{A(\mathrm{GT})}^{(j)} \\ \text{C-Rec}(\boldsymbol{X}_I) = \frac{1}{|\boldsymbol{X}_{A(\mathrm{GT})}|} \sum_{\boldsymbol{X}_{A(\mathrm{GT})}^{(j)} \in \boldsymbol{X}_{A(\mathrm{GT})}} \max_{\boldsymbol{X}_A^{(i)} \in \boldsymbol{X}_A} \boldsymbol{X}_A^{(i)} \cdot \boldsymbol{X}_{A(\mathrm{GT})}^{(j)} \end{cases} \tag{22}$$

where $\boldsymbol{X}_{A(\text{GT})}$ denotes the action scheduling ground truth in which $\boldsymbol{X}_{A(\text{GT})}^{(j)}$ is one of the expected actions. And ExactMatch evaluates the ratio of testing O&M instructions that necessarily and sufficiently fulfill their intended scheduling without further interventions. C-Rec is computed as Equation (23):

$$ExactMatch = n_{\text{Per}} / n_{\text{TC}} \tag{23}$$

where $n_{\text{Per}}$ denotes the number of testing O&M instructions that necessarily and sufficiently fulfill their intended scheduling without further interventions. $n_{\text{TC}}$ denotes the total number of testing composite O&M instructions, i.e., 1032 in this study.

Average intervention count (Itv) measures the frequency of manual intervention required for the agent to parse and execute composite instructions. It is expressed as the average number of additional user intervention rounds needed per 100 composite instructions to achieve its intended purpose, as formalized in Equation (24):

$$\text{Itv} = \frac{100}{n} \sum_{i=1}^{n} rd^{(i)} \tag{24}$$

where $n$ is the number of composite instructions. $rd^{(i)}$ denotes the number of additional instruction rounds required to fulfill the intended purpose of composite instruction $i$.

Average relative task time (R-Time) examines the impact of each agent module on composite task execution efficiency. It is defined as the ratio of the average execution time of composite instructions that can be successfully executed with zero intervention under a given architecture to the execution time under the complete BOMMA architecture, as shown in Equation (25):

$$\text{R-Time} = \frac{\frac{1}{n_{\text{Inc}}} \sum_{i=1}^{n_{\text{Inc}}} tm'^{(i)}}{\frac{1}{n_{\text{Com}}} \sum_{j=1}^{n_{\text{Com}}} tm^{(j)}} \tag{25}$$

where $n_{\text{Inc}}$, $n_{\text{Com}}$ respectively denote the number of zero-intervention composite instructions achieving their purpose under types of incomplete BOMMA architecture or the complete one, $tm'^{(i)}$, $tm^{(j)}$ respectively represent the time taken to complete each instruction under the two architectures.

Finally, the three variants of BERT-Score [46] are employed to separately assess the semantic similarity between the O&M action reasoning generated by BOMMA and manual O&M reasoning written by human experts (the authors of the O&M actions). This comparison verifies the enhancement of the physical-mechanical mechanism on the agent's O&M interpretation capability, as formalized in Equation (26):

$$\begin{cases} R_{BERT} = \frac{1}{|\boldsymbol{X}|} \sum_{x^{(i)} \in \boldsymbol{X}} \max_{x_{\mathrm{GT}}^{(j)} \in \boldsymbol{X}_{\mathrm{GT}}} x^{(i)} \cdot x_{\mathrm{GT}}^{(j)} \\ P_{BERT} = \frac{1}{|\boldsymbol{X}_{\mathrm{GT}}|} \sum_{x_{\mathrm{GT}}^{(j)} \in \boldsymbol{X}_{\mathrm{GT}}} \max_{x^{(i)} \in \boldsymbol{X}} x^{(i)} \cdot x_{\mathrm{GT}}^{(j)} \\ F1_{BERT} = 2 \cdot \frac{P_{BERT} \cdot R_{BERT}}{P_{BERT} + R_{BERT}} \end{cases} \tag{26}$$

where $R_{BERT}$, $P_{BERT}$, $F1_{BERT}$ denote BERT recall, precision, and F1-score, respectively, $\boldsymbol{X}_{\mathrm{GT}}$ denotes the token sequences of the manual reasoning, with $x_{\mathrm{GT}}^{(j)}$ representing an individual token in the sequences.

It is worth noting that, for all the above metrics, an action is considered successfully scheduled only if both the action scheduling and its parameterization match the query requirements and the action results are successfully obtained.

### 3.3 Result and Analysis

Based on the evaluation metrics and experimental settings described above, this section systematically analyses the performance of the BOMMA framework in multi-modal, multi-agent bridge O&M tasks. The experiments assess the agent's automatic action construction capability from the perspective of action scheduling accuracy, and further validate the contributions of the memory module, the multi-agent coordination mechanism, and the model's reasoning capacity to overall performance through ablation studies. Quantitative evaluation results are presented in Table 5, Figure 2 and Figure 3, while Figure 4 to Figure 7 illustrate various representative application scenarios of BOMMA. The following subsections elaborate on and analyze the phenomena observed in these experimental results.

Table 5 Simple Instructions Recall Rate for BOMMA's Automatically Constructed Actions

| Action | Simple Instruction | Correct Count (↑) | Rec (%, ↑) |
|---|---|---|---|
| DQA | 626 | 544 | 86.90 |
| VEH | 242 | 141 | 58.26 |
| VIS | 242 | 242 | **100.00** |
| VDD | 242 | 240 | 99.17 |
| STG | 242 | 141 | 58.26 |
| AEW | 626 | 553 | 88.34 |
| KMD | 240 | 239 | 99.58 |
| Total | 2460 | 2100 | 85.37 |

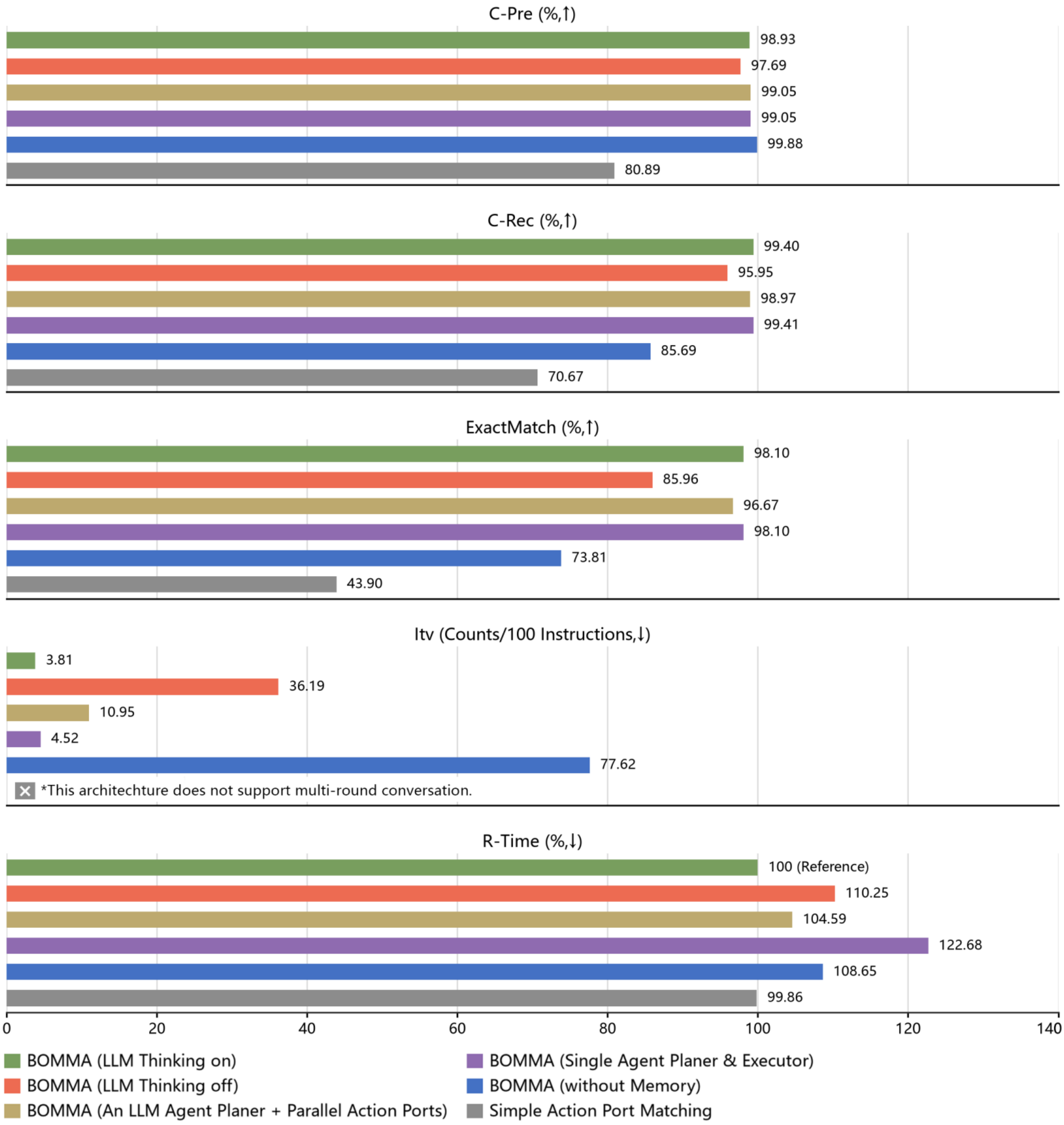

Figure 2 Performance of Different Agent Architectures on Composite Instructions

BOMMA shows a promising capability of nature language O&M instruction understanding and task execution, which significantly enhances the automation level of bridge O&M disposal. Table 5 shows that the scheduling recall rate of bridge O&M actions automatically constructed by BOMMA via the action creator is generally high, indicating the system is able to interpret simple instructions into O&M suitable actions and self-expand. And Figure 2 shows that BOMMA reaches a high-level accuracy of composite scheduling, indicating it is capable of complex instruction understanding and decomposing. Yet experiments exhibit differentiation that correlates with the semantics of action names. For actions with clearly defined functional boundaries—such as VIS and KMD—the recall rates approach near-perfection. However, certain O&M actions, notably VEH and STG, exhibit considerably lower scheduling accuracy, experiments reveal that simple instructions from Source III almost never successfully invoke these actions, suggesting that the underlying issue is the insufficient semantic association between action names and their physical-mechanical principles. Specifically, when the action name fails to intuitively reflect its inner principles which decipher the modalities (e.g., video streams, time-series signal arrays) and key parameters (e.g., spatial measurement points, time windows, threshold conditions), users of Source III—who lack understanding of the O&M actions and their internal physical-mechanical logic—tend to issue minimalist instructions that lack sufficient keywords. This prevents the agent from establishing effective semantic matching within the action space; even when the action is correctly identified, additional human intervention is often required for information completion.

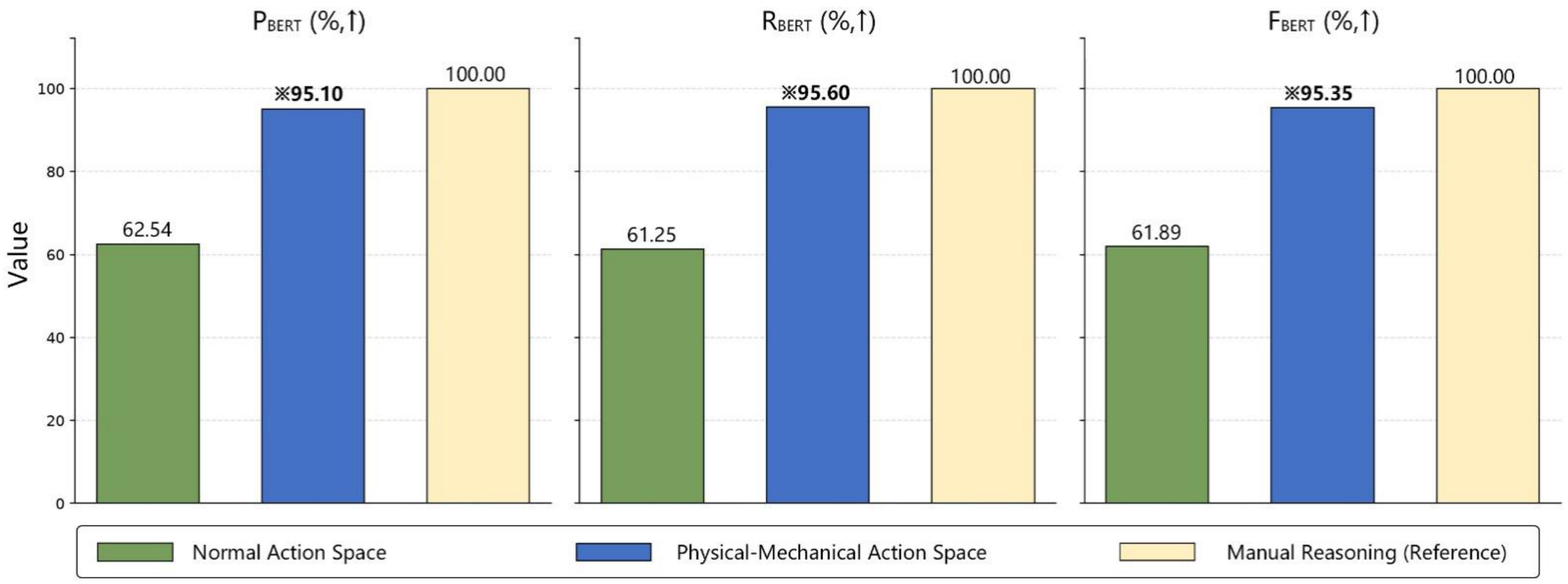

Figure 3 BOMMA Reasoning Results under Different Action Spaces

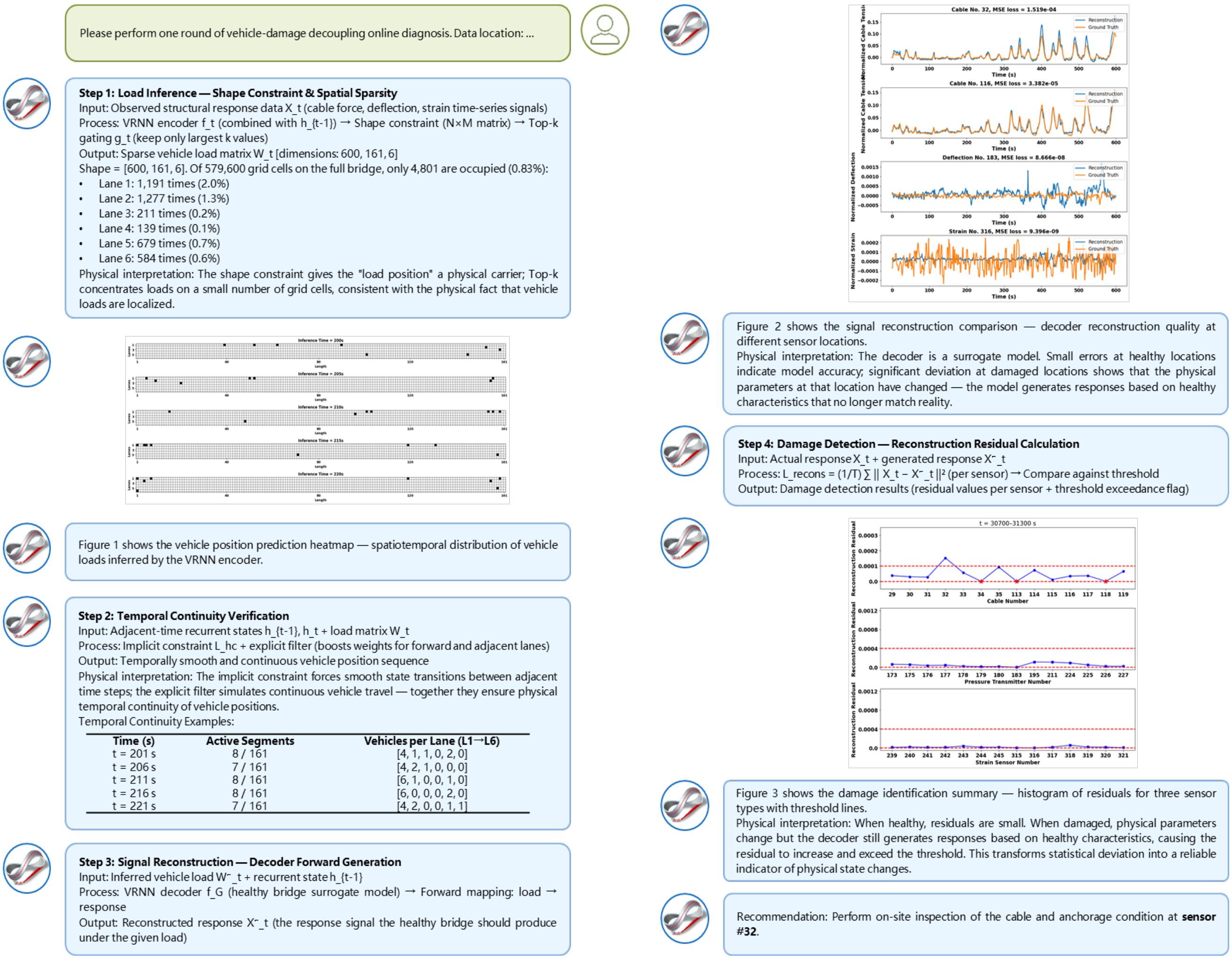

Figure 4 Example of Action Reasoning within the Physical-Mechanical Action Space

BOMMA is able to understand and interpret the agent's execution and action output, providing high confidence O&M automatic disposal. According to Figure 3, the physical-mechanical action space significantly enhances the agent's action reasoning response, which indicates that BOMMA is capable of parsing and explaining bridge O&M tasks. Figure 4 presents an example of O&M action reasoning of VDD generated with BOMMA's physical-mechanical action space, which successfully interprets the action execution into a comprehensive process and the underlying principles, including the process of vehicle position prediction for decoupling, temporal continuity verification for evaluation, decoupled signal reconstruction for comparison, and the final damage identification proposal, which shows high semantic consistency with expert-written explanations while providing a more explicit interpretation of the internal action process. And Figure 6 presents an example of BOMMA parsing the action's internal mechanics, in which it obtains the desired outcome of the instruction through the action's intermediate reasoning process. In contrast, the normal action space—while still capable of correct action scheduling and result retrieval—produces outputs that merely restate the execution outcomes without deeper interpretive reasoning. It is illustrated that physical-mechanical action space is a significant module for BOMMA to deliver comprehensive O&M execution and analysis.

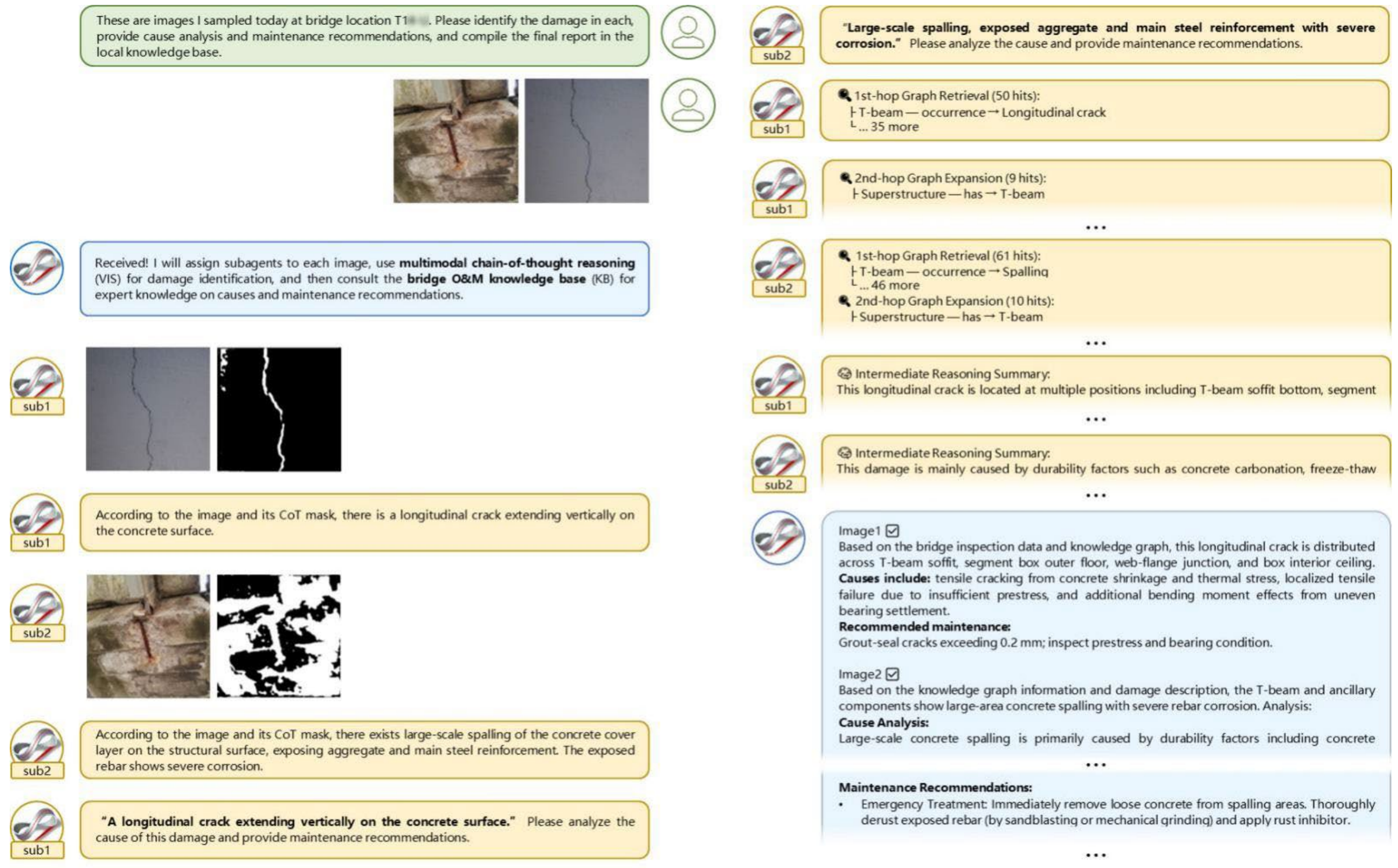

Figure 5 Expert Knowledge Trajectory Provided by Structured Knowledge Base

BOMMA enables the dynamic accumulation and solidification of memories in long-term working relationships with O&M users, providing more efficient and stable O&M interaction. Figure 2 illustrates that the removal of the memory module causes the most significant performance degradation. And experiments further observe that, for the 412 composite instructions derived from user protocols established by Source IV individuals (e.g., mapping "routine inspection" to a predefined combination of actions), removing the memory module reduces reasoning C-Pre, C-Rec, ExactMatch respectively by 9.72%, 51.07%, 84.05% and necessitates additional 361 rounds of human interventions—accounting for the predominant share of the overall performance drop. It aligns with expectations since the condition $\boldsymbol{X}_M$ is lost in Equation (1). Therefore the first step of the BOMMA close-loop, i.e.,

O&M task understanding and action scheduling shown in Equation (2), are forced to be reasoned in the absence of the prior memory tokens, consequently leading to impairing the agent's ability to correctly interpret user intent.

BOMMA's multi-agent coordination system significantly improves the execution efficiency of composite O&M instructions. Figure 2 illustrates that, compared to the single-agent planner-executor integrated architecture and the single-agent planner coordinated with non-agentic O&M action subprocesses, the complete BOMMA architecture (with multi-agent system) reduces the average time cost of testing composite O&M instructions while exerting negligible impact on O&M scheduling accuracy. This is because these two simpler architectures persistently occupy the main session, unable to distribute action execution and reflection to other reasoning-capable bodies, which forces concurrently executable O&M actions into serial processing and thereby forfeits the mobility advantage of the agent. This mechanism of BOMMA essentially transforms the serial O&M tasks into a distributed and parallel execution architecture. Persistent and computationally intensive actions, which are prevalent in bridge O&M such as time sequence and video data sampling and the action VEH, are offloaded to independent, instantized-on-demand, and concurrent sub-agents for concurrent disposal without serially occupying the core LLM's context window and reasoning time. The multi-agent system becomes particularly essential when a composite O&M instruction includes multiple concurrently executable O&M actions with significant disparities in time consumption. An example is presented in Figure 6 that the agent schedules VDD and VEH by physical-mechanical understanding, the former only takes a few minutes to finish while the latter needs 15 to 30 minutes, BOMMA is capable of detecting these two tasks have no dependency and executes them in parallel, consequently this O&M instruction is completed spending much less than the summation of both actions' duration, which indicates a prevention of long-duration actions propagating delays throughout the entire task chain provided by the multi-agent system. These results indicate that the primary contribution of the multi-agent organization is execution efficiency through parallel and distributed action handling, whereas scheduling accuracy is mainly determined by the planner, action-space representation, and contextual memory. It is worth noting that, the magnitude of parallel efficiency gain is contingent upon the agent's O&M task decomposition and planning capability, i.e., BOMMA must identify prerequisite dependencies and estimated execution times for each O&M action in order to schedule sub-processes appropriately. This, in turn, highly depends on the descriptions within the physical-mechanical action space.

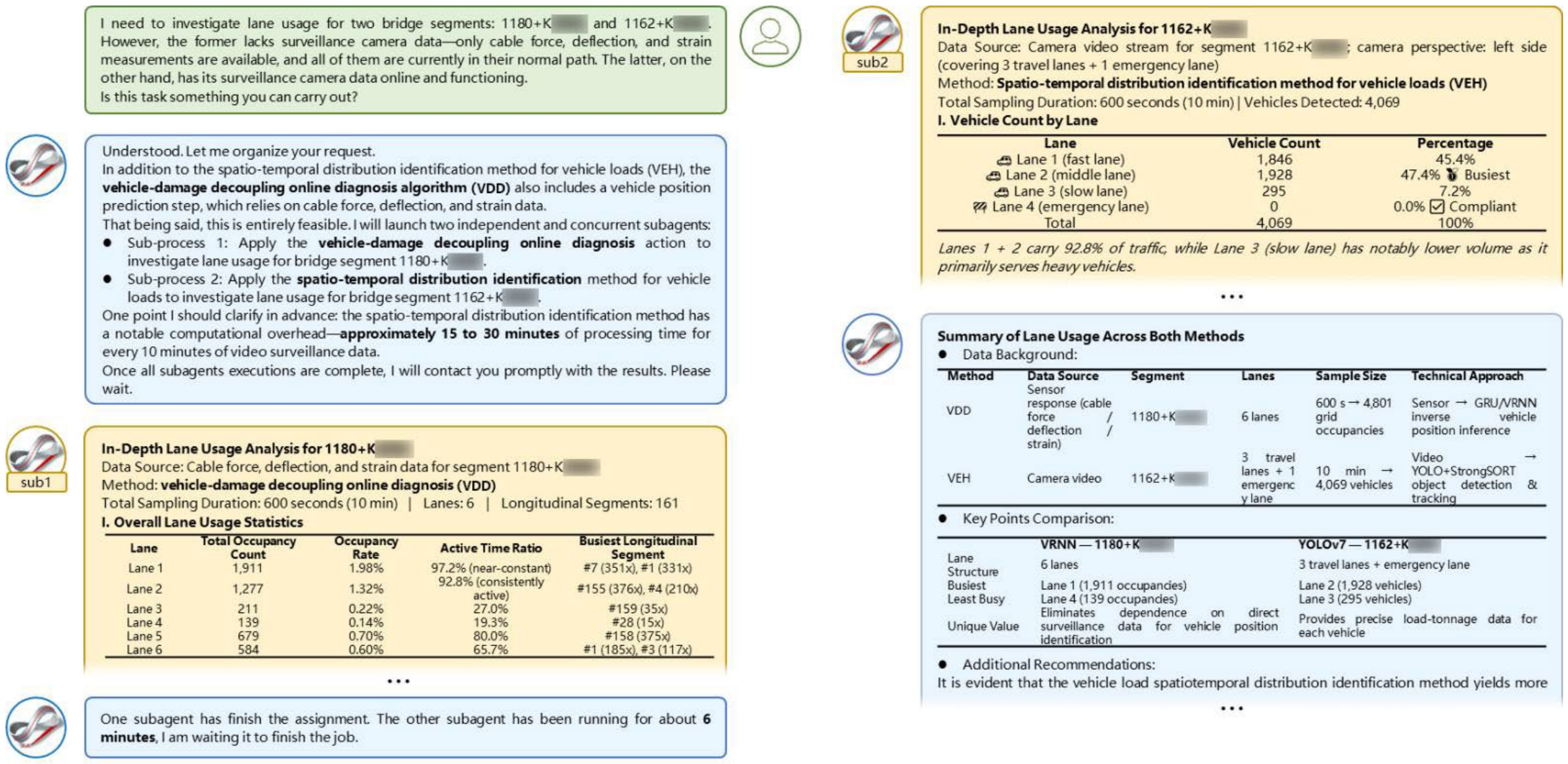

Figure 6 Example of Multi-Agent Usage of BOMMA

BOMMA is capable of proactively seeking user intervention when necessary, as illustrated in Figure 7, enabling stable O&M disposal. Experimental results indicate that the majority of multi-round human intervention needs is originate from Source III and Source IV individuals, which aligns with expectations on account of lacking physical-mechanical understanding of O&M actions and inherent memory reference, i.e., the nature of these interventions lies in information gaps encountered during reasoning. But different from general proposed solitary LLMs which favor resorting to generalized token prediction ability to fabricate answers with insufficient O&M instructions (such as lacking the structural parameters of AEW or the target response label of STG), a behavior that would lead to unsupported execution under incomplete instructions, the agent promptly acknowledges that the task cannot proceed due to O&M information deficiencies under the constraints of the physical-mechanical action space, feeds back the current problem and seeks supplementary input from users. This phenomenon demonstrates that BOMMA can accurately identify information gaps for O&M disposal.

Experiments show that simply providing the required information in natural language enables all initially failed O&M instructions to achieve their intended purposes, in other words, while this agentic mode entails a moderate trade-off in terms of automation O&M disposal level, it offers greater O&M application stability and is more economical in terms of token consumption and context length by reducing unsupported execution and reasoning tokens. For practical engineering domains such as bridge O&M that prioritize cost, quality, and safety, it represents a more stable and reliable mode.

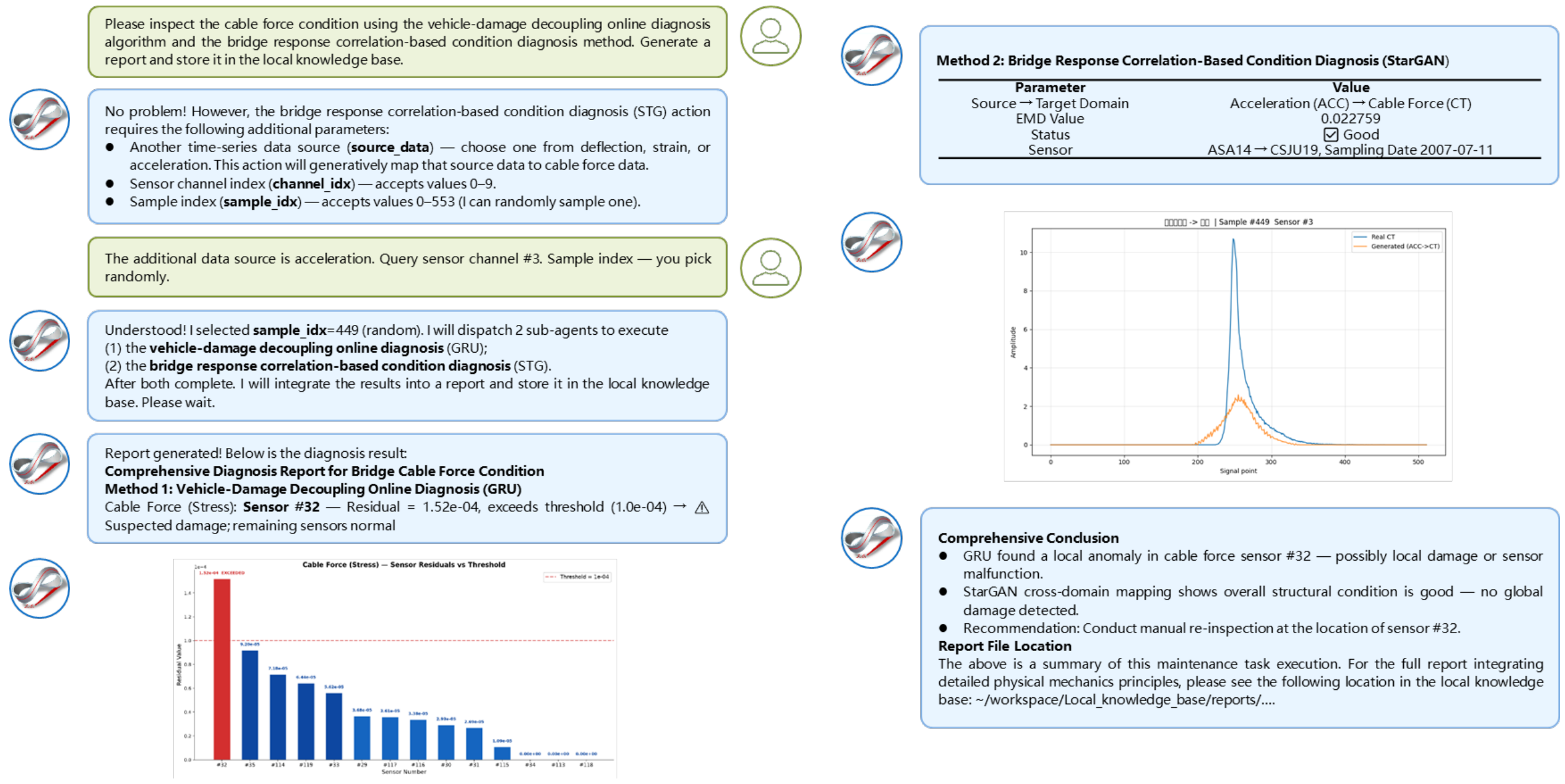

Figure 7 Example of BOMMA Proactively Seeking User Intervention

In summary, the performance of each module within the BOMMA framework can be characterized as follows: the physical-mechanical action space significantly enhances the reasoning ability for task execution; the effectiveness of the action creator is constrained by the semantic association density between action names and their trigger expressions; the memory module serves as the cornerstone for correctly understanding and decomposing composite instructions in long-term O&M scenarios; the multi-agent system enhances task efficiency through parallel execution without compromising decision quality; and the interactive agent workflow—through its information-gap identification and proactive intervention mechanism—provides essential safety boundaries and a foundation of trust for engineering-level deployment. Collectively, the experiments and their principle-based analyses presented above demonstrate the abilities of the proposed autonomy formalization methodology, validate the functional and architectural advantages of BOMMA, and offer clear guidance for subsequent optimization and practical deployment.

## 4 Conclusion

This study addresses the following research problems that constrain the intelligent transformation of bridge O&M. Heterogeneous data, methods, and algorithmic models remain fragmented, and existing workflows lack the autonomous coordination ability across composite O&M objectives. Historical tacit and experiential expert knowledge remain difficult to formalize, retrieve, and trace in support of O&M decisions.

To address these research problems, this study proposes a knowledge-enhanced autonomy formalization methodology for bridge O&M and applies it to construct BOMMA, a multi-modal, multi-agent bridge O&M system. First, the methodology formalizes bridge O&M autonomy as a staged coupling of physical-mechanical-grounded action representation, multi-agent organizational coordination, and long-term operational memory. The physical-mechanical action space grounds instruction comprehension and task decomposition; the multi-agent workflow coordinates action parallel execution and reflection; and long-term memory preserves protocols and operational context. Second, an expert O&M knowledge graph organizes tacit and empirical knowledge into evidence chains, thereby supporting traceable O&M decisions. Together, these two capabilities address the fragmentation of O&M resources and the limited traceability of expert knowledge in decision-making.

Experimental results verify that the proposed methodology makes contributions to the aforementioned problems. The autonomy is able to arrange heterogeneous O&M actions into coordination, that BOMMA achieves reliable action scheduling accuracy and complex task reasoning performance when processing bridge O&M instructions spanning diverse sources, categories, and data modalities, meanwhile the physical-mechanical action representations substantially strengthens the agent’s ability to interpret and implement bridge O&M tasks, the multi-agent workflow boosts overall task throughput via distributed computation while fully preserving reasoning quality, the memory unit reliably supports accurate parsing and feasible execution of composite instructions and pre-defined operational protocols. And O&M expert knowledge graph enables a traceable decision-making process.

Addition discoveries have been made in the experiments: the interactive workflow integrates proactive information gap identification and targeted intervention mechanisms, effectively mitigating the inherent hallucination tendency of the core LLM under insufficient input information and establishing critical safety guardrails for practical engineering-scale deployment. Collectively, these results validate the effectiveness and capability boundaries of the proposed bridge O&M specific autonomy, and clarify the distinct roles of the modules of BOMMA, confirming their potential to address critical challenges in the intelligent transformation of bridge O&M.

For future work, more bridge O&M actions should be aggregated and methodologies for coordinating the complexity will be urged [47]. In parallel, it is intended to investigate construction methodologies for all-time-domain, multi-scenario, lightweight bridge O&M agent clusters for the continuous and evolving health conditions of diverse bridge types over lifecycle [48].

## 5 Acknowledgement

This study was supported by Guizhou Provincial Major Science and Technology Special Project [Project No. Guizhou Major Science and Technology Contract (2025) 033], Guizhou Provincial Major Science and Technology Special Project [Project No. Guizhou Scientific Research Platform Contract (2025) 013], National Key Research and Development Program of China [Grant No. 2021YFF0501003], and Jinan City Science and Technology Program Project [Grant No. 202428059]. The financial support is greatly appreciated.

Special thanks to Qi Li, Teng Li, and Xinlei Zhang for their contributions about the construction and evaluation of the physical-mechanical action space.

## Appendix A Automatic O&M Action Creating Workflow of BOMMA

The proposed physical-mechanical action space comprises not only the full set of tools and skills (hereafter referred to collectively as actions) available to the agent, but also semantically embeds the physical-mechanical rationale underpinning each action. In effect, this action space directly defines both the agent's understanding of actions and the scope of its actionable capabilities. Leveraging the advantages of LLM-based few-shot and zero-shot task learning, BOMMA is capable of automatically constructing standardized actions tailored to user requirements. This, in turn, enables the seamless construction, integration, and expansion of the action space. The process comprises closed-loop steps: requirement source reading, standardized skill definition generation, test execution, and feedback-driven optimization.

Initially, the agent's action space contains only a set of basic actions—such as memory viewing, code compilation, web search, and action creation. The user informs the agent via natural language of the need to create a new action, specifying the action's content with either nature language or code. This query is then communicated to the action creator, as formalized in Equation (A.1):

$$p_{\text{Creator}} = p(\boldsymbol{X}_{\text{Creator}} | \boldsymbol{X}_I, \boldsymbol{X}_{\text{PM}}, \boldsymbol{X}_M) \tag{A.1}$$

where $p_{\text{Creator}}$ denotes the probability of BOMMA scheduling the action creator, $\boldsymbol{X}_{\text{Creator}}$ denotes the token sequence of the action creator.

Subsequently, guided by the action creator, the core LLM generates an initial version of the action in a standardized format, drawing upon the user instruction, the provided source code, and the underlying physical-mechanical rationale of the action (Table B.1 illustrates the detailed structure of a generated action). Equation (A.2) describes this step:

$$\begin{cases} p(\boldsymbol{X}_{\text{NA}}, \boldsymbol{X}_{\text{Creator}} | \boldsymbol{X}_I, \boldsymbol{X}_{\text{PM}}, \boldsymbol{X}_M) = p_{\text{NA}} \cdot p_{\text{Creator}} \\ p_{\text{NA}} = p(\boldsymbol{X}_{\text{NA}} | \boldsymbol{X}_{\text{Creator}}, \boldsymbol{X}_I, \boldsymbol{X}_{\text{PM}}, \boldsymbol{X}_M) \end{cases} \tag{A.2}$$

where $\boldsymbol{X}_{\text{NA}}$ denotes the token sequence of the initial new action, and $p_{\text{NA}}$ denotes the generation probability of $\boldsymbol{X}_{\text{NA}}$.

Thereafter, under the guidance of the action creator, BOMMA proceeds to test and optimize the

action—a process that encompasses both action execution optimizer and action description optimizer—by guiding the core LLM to generate a set of queries according to the new action. as illustrated in Equations (A.3):

$$\begin{cases} p(\boldsymbol{X}_{\mathrm{OPR}}, \boldsymbol{X}_{\mathrm{NA}}, \boldsymbol{X}_{\mathrm{Creator}} | \boldsymbol{X}_{I}, \boldsymbol{X}_{\mathrm{PM}}, \boldsymbol{X}_{M}) = p_{\mathrm{OPR}} \cdot p_{\mathrm{NA}} \cdot p_{\mathrm{Creator}} \\ p_{\mathrm{OPR}} = p(\boldsymbol{X}_{\mathrm{OPR}} | \boldsymbol{X}_{\mathrm{NA}}, \boldsymbol{X}_{\mathrm{Creator}}, \boldsymbol{X}_{I}, \boldsymbol{X}_{\mathrm{PM}}, \boldsymbol{X}_{M}) \\ \boldsymbol{X}_{\mathrm{OPR}} = \{\boldsymbol{X}_{Q(\alpha)}, \boldsymbol{X}_{Q(\beta 1)}, \boldsymbol{X}_{Q(\beta 2)}\} \end{cases} \tag{A.3}$$

where $\boldsymbol{X}_{\mathrm{OPR}}$ denotes the token sequence of testing queries generated by optimizers, in which $\boldsymbol{X}_{Q(\alpha)}, \boldsymbol{X}_{Q(\beta 1)}, \boldsymbol{X}_{Q(\beta 2)}$ respectively denote the token sequences of execution testing query set, query set supposed to trigger the new action, query set not supposed to trigger the new action. $p_{\mathrm{OPR}}$ denotes the conditional probability distribution of $\boldsymbol{X}_{\mathrm{OPR}}$. Note that $\boldsymbol{X}_{Q(\alpha)}, \boldsymbol{X}_{Q(\beta 1)}, \boldsymbol{X}_{Q(\beta 2)}$ are mutually independent, which is resulting in the following Equation (A.4):

$$\begin{cases} p_{\mathrm{OPR}} = p_{\mathrm{Q}(\alpha)} \cdot p_{\mathrm{Q}(\beta 1)} \cdot p_{\mathrm{Q}(\beta 2)} \\ p_{\mathrm{Q}(\alpha)} = p\left(\boldsymbol{X}_{\mathrm{Q}(\alpha)} | \boldsymbol{X}_{\mathrm{NA}}, \boldsymbol{X}_{\mathrm{Creator}}, \boldsymbol{X}_{I}, \boldsymbol{X}_{\mathrm{PM}}, \boldsymbol{X}_{M}\right) \\ p_{\mathrm{Q}(\beta 1)} = p\left(\boldsymbol{X}_{\mathrm{Q}(\beta 1)} | \boldsymbol{X}_{\mathrm{NA}}, \boldsymbol{X}_{\mathrm{Creator}}, \boldsymbol{X}_{I}, \boldsymbol{X}_{\mathrm{PM}}, \boldsymbol{X}_{M}\right) \\ p_{\mathrm{Q}(\beta 2)} = p\left(\boldsymbol{X}_{\mathrm{Q}(\beta 2)} | \boldsymbol{X}_{\mathrm{NA}}, \boldsymbol{X}_{\mathrm{Creator}}, \boldsymbol{X}_{I}, \boldsymbol{X}_{\mathrm{PM}}, \boldsymbol{X}_{M}\right) \end{cases} \tag{A.4}$$

where $p_{\mathrm{Q}(\alpha)}, p_{\mathrm{Q}(\beta 1)}, p_{\mathrm{Q}(\beta 2)}$ respectively denote the conditional probability distributions of $\boldsymbol{X}_{Q(\alpha)}, \boldsymbol{X}_{Q(\beta 1)}, \boldsymbol{X}_{Q(\beta 2)}$. According to $\boldsymbol{X}_{Q(\alpha)}$, BOMMA tests the execution with the new action space (containing the new action) or old action space(not containing the new action), illustrated in Equation (A.5):

$$\begin{cases} p\left(\boldsymbol{X}_{\mathrm{NO}}, \boldsymbol{X}_{\mathrm{OO}}, \boldsymbol{X}_{Q(\alpha)}, \boldsymbol{X}_{\mathrm{NA}}, \boldsymbol{X}_{\mathrm{Creator}} | \boldsymbol{X}_{I}, \boldsymbol{X}_{\mathrm{PM}}, \boldsymbol{X}_{M}\right) = p_{\mathrm{NOOO}} \cdot p_{\mathrm{Q}(\alpha)} \\ p_{\mathrm{NOOO}} = p\left(\boldsymbol{X}_{\mathrm{NO}}, \boldsymbol{X}_{\mathrm{OO}} | \boldsymbol{X}_{Q(\alpha)}, \boldsymbol{X}_{\mathrm{NA}}, \boldsymbol{X}_{\mathrm{Creator}}, \boldsymbol{X}_{I}, \boldsymbol{X}_{\mathrm{PM}}, \boldsymbol{X}_{M}\right) \end{cases} \tag{A.5}$$

where $\boldsymbol{X}_{\mathrm{NO}}, \boldsymbol{X}_{\mathrm{OO}}$ respectively denote the token sequences of new action space execution output and old action space execution output. $p_{\mathrm{NOOO}}$ denotes the conditional probability distribution of $\boldsymbol{X}_{\mathrm{NO}}$ and $\boldsymbol{X}_{\mathrm{OO}}$. Analogously, according to $\boldsymbol{X}_{Q(\beta 1)}, \boldsymbol{X}_{Q(\beta 2)}$, the corresponding action scheduling are reasoned, illustrated in Equation (A.6) and (A.7):

$$\begin{cases} p\left(\boldsymbol{X}_{A(\beta 1)}, \boldsymbol{X}_{Q(\beta 1)}, \boldsymbol{X}_{\mathrm{NA}}, \boldsymbol{X}_{\mathrm{Creator}} | \boldsymbol{X}_{I}, \boldsymbol{X}_{\mathrm{PM}}, \boldsymbol{X}_{M}\right) = p_{A(\beta 1)} \cdot p_{Q(\beta 1)} \\ p_{A(\beta 1)} = p\left(\boldsymbol{X}_{A(\beta 1)} | \boldsymbol{X}_{Q(\beta 1)}, \boldsymbol{X}_{\mathrm{NA}}, \boldsymbol{X}_{\mathrm{Creator}}, \boldsymbol{X}_{I}, \boldsymbol{X}_{\mathrm{PM}}, \boldsymbol{X}_{M}\right) \end{cases} \tag{A.6}$$

$$\begin{cases} p\left(\boldsymbol{X}_{A(\beta 2)}, \boldsymbol{X}_{Q(\beta 2)}, \boldsymbol{X}_{\mathrm{NA}}, \boldsymbol{X}_{\mathrm{Creator}} \middle| \boldsymbol{X}_I, \boldsymbol{X}_{\mathrm{PM}}, \boldsymbol{X}_M\right) = p_{A(\beta 2)} \cdot p_{Q(\beta 2)} \\ p_{A(\beta 2)} = p\left(\boldsymbol{X}_{A(\beta 2)} \middle| \boldsymbol{X}_{Q(\beta 2)}, \boldsymbol{X}_{\mathrm{NA}}, \boldsymbol{X}_{\mathrm{Creator}}, \boldsymbol{X}_I, \boldsymbol{X}_{\mathrm{PM}}, \boldsymbol{X}_M\right) \end{cases} \tag{A.7}$$

where $\boldsymbol{X}_{A(\beta 1)}, \boldsymbol{X}_{A(\beta 2)}$ respectively denote the action scheduling results of query set $\boldsymbol{X}_{Q(\beta 1)}, \boldsymbol{X}_{Q(\beta 2)}$, and $p_{A(\beta 1)}, p_{A(\beta 2)}$ respectively denote the conditional probability distributions of $\boldsymbol{X}_{A(\beta 1)}, \boldsymbol{X}_{A(\beta 2)}$. Note that $\boldsymbol{X}_{\mathrm{NO}}$, $\boldsymbol{X}_{\mathrm{OO}}$, $\boldsymbol{X}_{A(\beta 1)}$ and $\boldsymbol{X}_{A(\beta 2)}$ are materials for optimizing the initial new action, which are mutually independent.

Finally, the optimization materials are consolidated to update $\boldsymbol{X}_{\mathrm{NA}}$, yielding the deployable new action, as expressed in Equation (A.8):

$$\begin{cases} p(\boldsymbol{X}_{\mathrm{NA}}^{*}, \boldsymbol{X}_{\mathrm{OPT}}, \boldsymbol{X}_{\mathrm{OPR}}, \boldsymbol{X}_{\mathrm{NA}}, \boldsymbol{X}_{\mathrm{Creator}} | \boldsymbol{X}_I, \boldsymbol{X}_{\mathrm{PM}}, \boldsymbol{X}_M) \\ = p_{\mathrm{NA}*} \cdot p_{\mathrm{OPT}} \cdot p_{\mathrm{OPR}} \cdot p_{\mathrm{NA}} \cdot p_{\mathrm{Creator}} \\ p_{\mathrm{NA}*} = p(\boldsymbol{X}_{\mathrm{NA}}^{*} | \boldsymbol{X}_{\mathrm{OPT}}, \boldsymbol{X}_{\mathrm{OPR}}, \boldsymbol{X}_{\mathrm{NA}}, \boldsymbol{X}_{\mathrm{Creator}}, \boldsymbol{X}_I, \boldsymbol{X}_{\mathrm{PM}}, \boldsymbol{X}_M) \\ p_{\mathrm{OPT}} = p(\boldsymbol{X}_{\mathrm{OPT}} | \boldsymbol{X}_{\mathrm{OPR}}, \boldsymbol{X}_{\mathrm{NA}}, \boldsymbol{X}_{\mathrm{Creator}}, \boldsymbol{X}_I, \boldsymbol{X}_{\mathrm{PM}}, \boldsymbol{X}_M) \\ \boldsymbol{X}_{\mathrm{OPR}} = \left\{\boldsymbol{X}_{A(\beta 1)}, \boldsymbol{X}_{A(\beta 2)}, \boldsymbol{X}_{\mathrm{NO}}, \boldsymbol{X}_{\mathrm{OO}}\right\} \end{cases} \tag{A.8}$$

where $\boldsymbol{X}_{\mathrm{NA}}^{*}$ denotes the token sequence of the deployable new action. $\boldsymbol{X}_{\mathrm{OPT}}$ denotes the aggregated optimization material set. $p_{\mathrm{NA}*}, p_{\mathrm{OPT}}$ respectively denote the conditional probability distributions of $\boldsymbol{X}_{\mathrm{NA}}^{*}$ and $\boldsymbol{X}_{\mathrm{OPT}}$. Due to the fact that the optimization materials are mutually independent, the following Equation (A.9) is formalized:

$$p_{\mathrm{OPT}} = p_{A(\beta 1)} \cdot p_{A(\beta 2)} \cdot p_{\mathrm{NOOO}} \tag{A.9}$$

In summary, the following Equation (A.10) results:

$$\begin{aligned} & p(\boldsymbol{X}_{\mathrm{NA}}^{*}, \boldsymbol{X}_{\mathrm{OPT}}, \boldsymbol{X}_{\mathrm{OPR}}, \boldsymbol{X}_{\mathrm{NA}}, \boldsymbol{X}_{\mathrm{Creator}} | \boldsymbol{X}_{\mathrm{I}}, \boldsymbol{X}_{\mathrm{PM}}, \boldsymbol{X}_{\mathrm{M}}) \\ & = p_{\mathrm{NA}*} \cdot p_{A(\beta 1)} \cdot p_{A(\beta 2)} \cdot p_{\mathrm{NOOO}} \cdot p_{\mathrm{Q}(\alpha)} \cdot p_{\mathrm{Q}(\beta 1)} \cdot p_{\mathrm{Q}(\beta 2)} \cdot p_{\mathrm{NA}} \cdot p_{\mathrm{Creator}} \end{aligned} \tag{A.10}$$

The aforementioned process can be iterated to perform multiple rounds of action optimization, enabling progressive refinement.

## Appendix B Composition of the Knowledge-Enhanced Autonomy for Bridge O&M

Table B.1 Composition of an Action in the Physical-Mechanical Action Space

| Content | Necessity | Connotation |
| --- | --- | --- |
| Abstract | Necessary | Name, brief introduction, workflow of the action. |
| Physical & mechanical principle | Necessary | The detailed physical and mechanical principles of the O&M action. |
| Parameters | Necessary | All the required inputs of the action and their names and functions. |
| Reference | Optional | The example queries to trigger the action;<br>The standardized input example(s) of the action;<br>Scheduling for actions other than itself;<br>Other references urged but cannot fit in the Abstract;<br>etc. |
| Assets | Optional | Data, executable script, output history, etc. of the action. |
| Others | Optional | Other customized content needed for the action. |

Table B.2 Composition of BOMMA's Memory

| Composition | Content |
|---|---|
| Basic information | Agent's identity, action guidelines, working directory, security protocols, etc. |
| Users' information | Users' identities, habits, agreements, other customized long-term memories. |
| Other customized content | Such as data path, website, etc. |

## Appendix C Examples for Each Type of Testing Instruction

Table C.1 Examples for Each Type of Testing Instruction

| Type | Source | Instruction | Label |
|---|---|---|---|
| Simple Instruction | Source I | Based on the bridge response correlation pattern, compare the correlation between the cable-force and acceleration time histories under normal conditions. If the correlation weakens or the amplitude or phase becomes abnormal, determine whether the structural condition at this location has changed. | STG |
| | Source II | I need full-field vehicle position information, but only static response data are currently available and no surveillance video is provided. Please attempt the task with the available data. | VDD |
| | Source III | Please check if early warning should be informed. | AEW |
| | Source IV | \ | \ |
| Composite Instruction | Source I | Please use the cable-force and deflection data to separate the vehicle-load response from the structural-damage response, compare the correlation patterns of the two responses, determine the structural condition, and record the conclusion in the O&M knowledge base with a timestamp. | STG, VDD, KMD |

| | | |
|---|---|---|
| Source II | Please inspect the current inspection image for signs of apparent damage such as cracks, spalling, or seepage. Then use the deflection and strain data to identify potential concealed damage, explain the basis for the judgment, and compile the results into an inspection record. | VIS, VDD, KMD |
| Source III | Please analyze all types of modal data collected this week and generate a weekly report. | DQA, VIS, VDD, STG , VEH, KMD |
| Source IV | Please conduct a routine inspection. | VIS, VDD, STG, VEH, KMD |

## Appendix D Subagent's Reflection Termination Conditions

The subagent operates within a reflective decision-making loop that systematically evaluates task execution outcomes against termination conditions, as illustrated in Figure D.1.

Firstly, upon receiving the initial execution results from its assigned actions, a subagent assesses whether direct evidence exists to address the sub-task query. If affirmative, it satisfies the "direct evidence" termination criteria; if negative, it invokes the secondary evaluation.

Secondly, a subagent assesses whether the assigned sub-task reaches the permitted number of trying loops. If affirmative, it satisfies the "max loop" termination criteria; if negative, it invokes the tertiary evaluation.

Thirdly, a subagent further assesses the current sub-task outputs. In cases where the current evidence is either weak or empty, the subagent explores alternative action choices to gather additional information. Should no viable alternative exist, the workflow satisfies the "no movement" termination criteria; else the subagent proceeds next loop of disposal.

Once a subagent meets one of the termination conditions, it accumulates all O&M executing results and possibly existing executing obstacles to generate the final O&M answer and feedback to users.

This design ensures that the subagent autonomously balances thoroughness with computational efficiency, preventing infinite loops while maximizing the utility of available evidence through multi-round reasoning and adaptive action rescheduling.

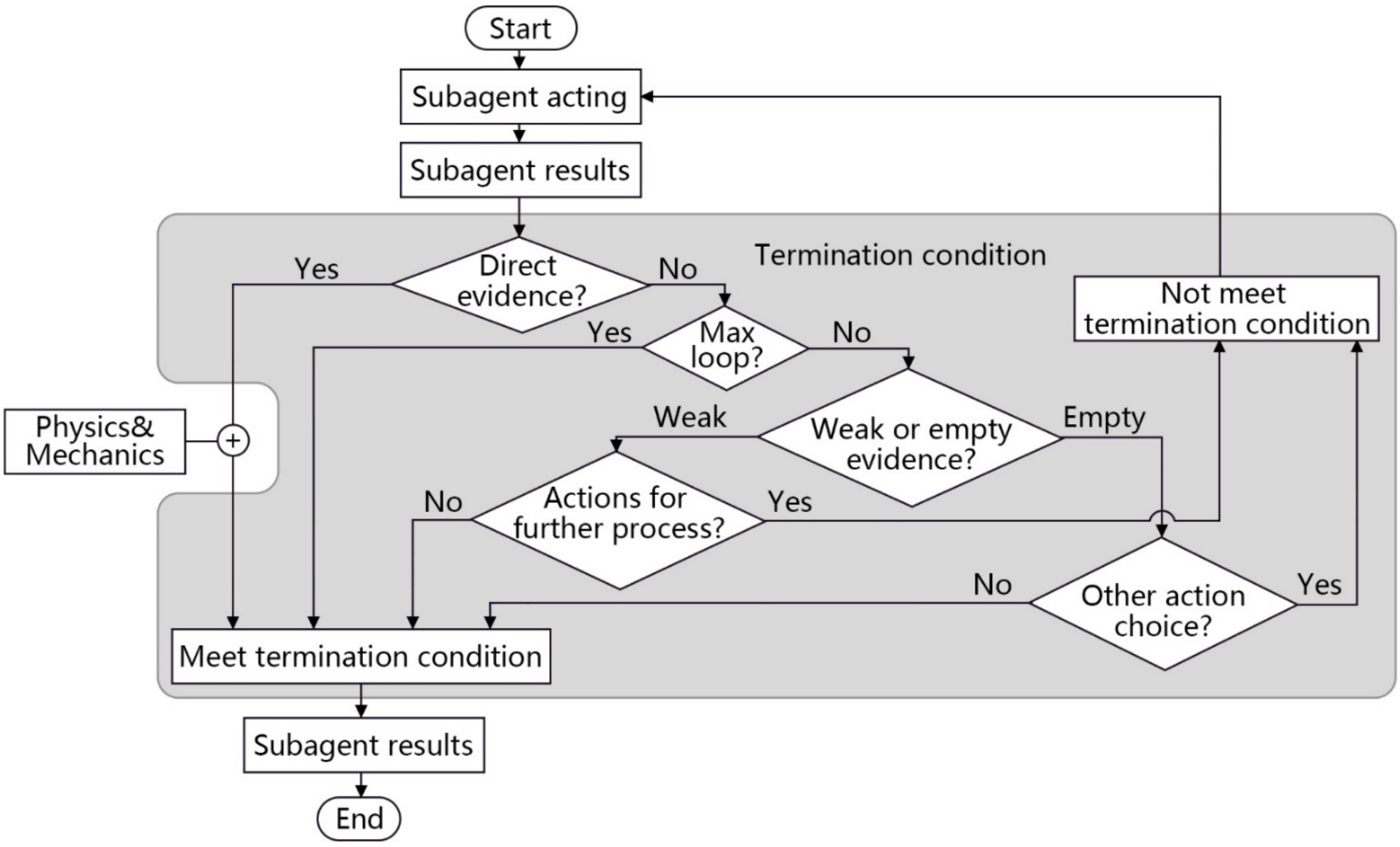

Figure D.1 Subagent Reflection Process